\documentclass[journal]{new-aiaa}
\usepackage{new-mysymb}
\usepackage{longtable}
\usepackage{multirow}
\usepackage{pbox}
\usepackage[hyphens]{url}
\usepackage{xcolor}
\usepackage{soul}
\usepackage{nomencl}
\usepackage{booktabs}
\usepackage{etoolbox}
\usepackage{adjustbox}
\usepackage[ruled,vlined,linesnumbered]{algorithm2e}

\newcommand\cK{\mathcal{K}}

\newcommand\bR{\mathbb{R}}

\newcommand\cD{\mathcal{D}}
\newcommand\cF{\mathcal{F}}

\newcommand\cL{\mathcal{L}}

\newcommand\cP{\mathcal{P}}

\newcommand\cU{\mathcal{U}}

\newcommand\cY{\mathcal{Y}}

\newcommand\Dt{\Delta t}

\soulregister\cite7
\soulregister\ref7

\newcommand\bP{\mathbb{P}}
\newcommand\bX{\mathbb{X}}
\newcommand\bY{\mathbb{Y}}
\newcommand\bC{\mathbb{C}}
\newcommand\koop{\cK_t}
\newcommand\vtvf{\boldsymbol{\varphi}}

\usepackage[utf8]{inputenc}
\usepackage{makecell}
\usepackage{float}
\usepackage{array}
\usepackage{graphicx}
\usepackage[version=4]{mhchem}
\usepackage{longtable,tabularx}
\usepackage{amsmath}
\usepackage{todonotes}

\title{Global Description of Flutter Dynamics via Koopman Theory}

\author{Jiwoo Song\footnote{Graduate Student Research Assistant, Student Member AIAA, jzs6565@psu.edu.} and Daning Huang\footnote{Assistant Professor, Member AIAA, daning@psu.edu.}}
\affil{Department of Aerospace Engineering, The Pennsylvania State University, University Park, PA, 16802}

\begin{document}
\footnotetext{Presented as Paper 2025-0423 at the AIAA SCITECH 2025 Forum, Orlando, Florida, January 6-10, 2025.}
\maketitle

\begin{abstract}
    This paper presents a novel parametrization approach for aeroelastic systems utilizing Koopman theory, specifically leveraging the Koopman Bilinear Form (KBF) model.  To address the limitations of linear parametric dependence in the KBF model, we introduce the Extended KBF (EKBF) model, which enables a global linear representation of aeroelastic dynamics while capturing stronger nonlinear dependence on, e.g., the flutter parameter.  The effectiveness of the proposed methodology is demonstrated through two case studies: a 2D academic example and a panel flutter problem.  Results show that EKBF effectively interpolates and extrapolates principal eigenvalues, capturing flutter mechanisms, and accurately predicting the flutter boundary even when the data is corrupted by noise.  Furthermore,  parameterized isostable and isochron identified by EKBF provides valuable insights into the nonlinear flutter system.
\end{abstract}

\section*{Nomenclature}

{\renewcommand\arraystretch{1.0}
\noindent\begin{longtable*}{@{}l @{\quad=\quad} l@{}}
$\vA$, $\vB$, $\vD$          & System matrices of linear dynamical systems\\
$\vf$, $\vg$                 & Dynamical system functions \\
$\koop$                      & Koopman operator\\
$\cL_{\vf}$, $\cL_{\vg}$     & Lie derivatives \\
$\vz$                        & Lifted coordinates \\
$\vtb,\beta_i$               & System parameter \\
$\vtL$, $\lambda$            & Eigenvalues \\ 
$\vtvf, \varphi_i$           & Eigenfunctions, $i$th eigenfunction \\
$\tilde{\vtf}(\cdot)$,$\tilde{\vty}(\cdot)$  & Mapping for lifted coordinates\\
$\vtU$, $\vtY$               & Matrices of right and left eigenvectors\\
$\Omega$                     & System parameter of panel flutter equation
\end{longtable*}}
% \todo{acronyms as well}

\section{Introduction}

Aeroelastic flutter is a dynamic, potentially catastrophic phenomenon in which aerodynamic forces interact with the dynamics of an aircraft or structure and lead to self-sustained oscillations \cite{Dowell2021}.
Many of the established flutter analysis tools rely on the assumption of linearity, as such methods are based on a local linearization at equilibrium points under steady flight conditions \cite{Jonsson2019}. However, the aeroelastic system is inherently high-dimensional and nonlinear due to the strong coupling between aerodynamics, structural dynamics, and sometimes rigid-body dynamics. This reliance on linear models, and hence the utility of such algorithms, may become problematic when dealing with stronger nonlinearities; this can be the cases for relatively unconventional aircraft configurations, such as very flexible aircraft \cite{guimaraes2023flexible}, morphing aircraft \cite{preston2025multibody}, and urban air mobility (UAM) \cite{su2022modeling}.
In these configurations, the aircraft may experience dynamics far away from equilibrium points, e.g., during drastic and agile maneuvers or complex gust excitations, so that flutter analysis tools based on local linearization might be no longer applicable.

This study aims to explore the possibility of applying global linearization techniques, such as the Koopman theory \cite{mauroy2020koopman}, for nonlinear flutter analysis.
Here, global linearization means constructing a linear model whose dynamics can exactly reproduce the observations of a nonlinear dynamical system within a finite domain of the state space; this is fundamentally different from locally linearized models that are technically only exact in the infinitesimal region near the linearization point.
Hence, the global linearization has the potential to provide a relatively comprehensive stability characterization of nonlinear aeroelastic phenomena; this can open up new venues in the analysis and understanding of nonlinear aeroelastic applications.

Next, to provide a precise scope of the current study, the existing flutter analysis methods are briefly reviewed.
The existing methods can be classified into two types based on their formulations: model-based methods and model-free methods. The model-based methods utilize the mathematical representation of the system dynamics to predict flutter; examples include, time marching \cite{goura2001time}, eigenvalue analysis \cite{hassig1971approximate}, and state-velocity method \cite{riso2022estimating}.
The model-based methods are applicable when an accurate dynamics model is available.  However, such availability of model is not always practical, e.g., when the system is highly complex, or when only partial observation of system states is available; the latter case is typical for the wind tunnel tests and in-flight measurements.
Hence, in this study, the focus is on the model-free setting, which is arguably more versatile in practice.
The model-free methods, also referred to as output-based methods, do not rely on system's prior information (e.g., the model) but instead identify either characteristic equations of the flutter or the system itself and apply it to predict the flutter. A classical example is the autoregressive (AR) method to estimate the eigenvalues \cite{box2015time}. A more recent example include the bifurcation forecasting technique, which uses the concept of critical slowing down \cite{ghadami2016bifurcation,Riso2021}, and another example is the matrix pencil method \cite{kiviaho2019flutter}.

Nevertheless, as pointed out earlier, most of the methods mentioned above rely on local linearization and do not explicitly address the nonlinearity of the system; the only exception is the bifurcation forecasting method.
Built on the normal form theory of nonlinear dynamics, bifurcation forecasting uses a polynomial model to relate the decay rate of one single oscillatory mode to the flutter parameter and the modal amplitude, and fits such a polynomial from observed data; the roots of the polynomial is subsequently used to predict the flutter boundary and construct the bifurcation curve.
However, this method requires mode filtering and phase-fixing steps, which complicates the process and can generate error if, for example, the transient response is too short so that the data do not have sufficient local extrema in the transient regime, or the system exhibits frequency coalescence.  In the latter case, estimation of envelope function could be challenging as there are multiple modes contributing to the envelope.
% A further complication is inherent to all model-free methods, that the data noise may cause spectral pollution and result in inaccurate estimates of frequencies and/or decay rates, and thus inaccurate prediction of flutter boundary.
A global linearization technique may systematically capture all modes involved in flutter, and thus overcome the limitations of the above method.

More specifically, Koopman operator theory as a tool for global linearization has shown initial success in modeling and control of nonlinear and turbulent fluid flows \cite{taira2017modal, mezic2013analysis, rowley2009Spectral}, as well as generic control applications \cite{Proctor2016DMDc, Korda2018MPC, Tsolovikos2021SPDMDControl, mauroy2020koopman, bevanda2021koopman, Yu2023}. The dynamic mode decomposition (DMD) has been developed and used extensively to identify Koopman operator numerically from data and shown promises in analyzing nonlinear high-dimensional dynamics \cite{taira2017modal,brunton2021modern,colbrook2024rigorous}. One example of DMD for aeroelasticity is higher-order DMD (HODMD) used for flutter analysis \cite{Mendez2021}, which directly treats multiple dynamical modes.  However, HODMD is diagnostic instead of predictive.  It extracts out modal information from measurement data at given flutter parameters but cannot extrapolate for flutter prediction at new flutter parameters.

To achieve a predictive capability under the Koopman framework, one needs to consider a parametrized Koopman model, having the flutter parameter as an input.
Note that many aeroelastic systems can be formulated as an input-affine form that is linear with respect to the flutter parameter, e.g., dynamic pressure.  Such input-affine dynamics can be globally linearized into the so-called Koopman Bilinear Form (KBF) \cite{Goswami2022} over the entire attraction basin under mild assumptions.  For aeroelastic systems, the attraction basin practically covers a major portion of state space around the attractors, i.e., the equilibrium points and limit cycles, beyond typical local linearization regimes.  Hence a KBF model, with the flutter parameter as input, has the potential to represent the nonlinear flutter characteristics in the entire flight regime of interest.  In addition, the input form of KBF allows it to readily model the flutter characteristics with varying flutter parameters, such as gust.  Our prior work \cite{song2024parametrized} has performed a preliminary study of KBF model for aeroelastic analysis, but the model faced limitations in predictive capability as it is limited to problems having the input-affine form in the flutter parameter.

To overcome the input-affine nature of the KBF method, this study aims to develop an Extended KBF (EKBF) model with higher-order flutter parameters in the model. 
Leveraging the Koopman operator theory and its spectral analysis \cite{Mezic2005Spectral, mezic2013analysis, Brunton2016KoopmanInvariant, Brunton2022KoopmanTheory}, this study aims to exploit the aeroelastic measurement data to the maximum and produce a predictive nonlinear flutter model across a wide range of flutter parameters.
Specifically, our goals are to,
\begin{compactenum}
    \item Present the basic formulation for the Extended Koopman Bilinear Form model for parameterized nonlinear dynamics of Hopf bifurcation (i.e., flutter).
    \item Establish connections between Koopman theory and flutter analysis, particularly in characterizing both pre-flutter and post-flutter regimes.
    \item Demonstrate the method on two numerical examples, with benchmark against classical flutter prediction methods.
\end{compactenum}

The remainder of the paper is organized as follows. In Sec. \ref{sec1}, the mathematical formulation of extension of Koopman bilinear form and its connection to the local stability analysis in both pre-flutter and post-flutter regimes are discussed. In Sec. \ref{sec2}, we present the interpretation of eigenfunction of the system and  evaluates the effectiveness of the new method in capturing eigenvalues through interpolation and extrapolation.  In Sec. \ref{sec3}, we summarize the conclusions of the paper with potential future works.

\section{Mathematical Formulation}\label{sec1}

This section first presents a mathematical description of the global linearization of a nonlinear flutter model.  Furthermore, the section discusses the relation between the conventional locally linearized model and Koopman-based globally linearized model in the context of stability analysis.  Lastly, the numerical implementation for the proposed algorithm is presented.

\subsection{Global Linearization of Parametric Nonlinear Systems via Koopman Theory}\label{ssc_dat}

We first briefly present the standard KBF model for the global linearization of input-affine nonlinear dynamics, and then discuss its extension to general nonlinear dynamics.

\subsubsection{Koopman Bilinear Form}

Consider the following form of dynamics,
\begin{equation}\label{eqn:systemEqn_m}
    \dot{\vx}=\vf(\vx)+\sum_{i=1}^p\vg_i(\vx)\beta_i
\end{equation}
where $\vx\in\bX\subseteq\bR^r$ is the state vector, $\vtb = [\beta_1 , \beta_2, \cdots, \beta_p]$ are input parameters, such as the dynamic pressure, $\vf:\bX\to\bR^r$ describes the autonomous dynamics, and $\vg_i:\bX\to\bR^r$ describes the effects of the parameter $\beta_i$ on the dynamics, e.g., due to aerodynamic loads.

Based on Eq. \eqref{eqn:systemEqn_m}, one can resort to Koopman Bilinear Form (KBF) \cite{Goswami2022,Jiang2022} to globally bilinearize such system.
Koopman theory is a convenient tool for the analysis of nonlinear dynamical systems.  When the system is autonomous and has an isolated stable equilibrium point, a global linearization can be achieved in the entire attracting basin of the system \cite{Budisic2012}; a similar linearization can be achieved in the case of limit cycles.  With such linearization, classical linear system theory can be leveraged to characterize the nonlinear dynamics.%, such as the timescales.

\paragraph{Autonomous case}
To explain the characterization of nonlinear dynamics, the autonomous case \cite{Mauroy2016}, \ie, when $\vtb=\mathbf{0}$, is considered.  In this case, the system generates a flow $\vF_t(\vx_0)=\vx(t)$ from an initial condition $\vx_0$.  The continuous time Koopman operator $\koop:\cF\rightarrow\cF$ is an infinite-dimensional linear operator such that
$\koop z = z\circ\vF_t$ for all $z\in\cF$, where $z:\bX\rightarrow\bC$ is a complex-valued observable function of the state vector $\vx$, $\cF$ is the function space of all possible observables, and $\circ$ denotes function composition.
As a linear operator, $\koop$ admits eigenpairs $(\lambda,\varphi)$ such that
\begin{equation}\label{eqn:koopman-eigen}
    \koop\varphi = \varphi\circ\vF_t = e^{\lambda t}\varphi\:,
\end{equation}
where $\lambda\in\bC$ and $\varphi\in\cF$ are the Koopman eigenvalue and Koopman eigenfunction, respectively.

The infinitesimal generator of $\koop$ associated with $\vf$, referred to as the Koopman generator, is defined as $\cL_{\vf}=\lim_{t\rightarrow0}\frac{\koop-I}{t}$, where $I$ is the identity operator, and turns out to be the Lie derivative $\cL_\vf=\vf\cdot\nabla$, with eigenpair $(\lambda,\varphi)$,
\begin{equation}\label{eqn:lie_eqn}
    \dot{\varphi} = \cL_{\vf}\varphi = \lambda\varphi\:.
\end{equation}

\paragraph{Lattice structure of the Koopman eigenvalues} \label{koopman_lattice}  One interesting property of Koopman operator is that it generates more eigenfunctions from the known eigenfunctions. In continuous time setup, if there are two eigenpairs $(\lambda_1,\varphi_1)$ and $(\lambda_2,\varphi_2)$, then
\begin{equation}\begin{split}
    \koop (\varphi_1 \varphi_2) &= \frac{d}{dt} (\varphi_1\varphi_2) \\
    &= \dot{\varphi}_1 \varphi_2 + \varphi_1 \dot{\varphi}_2 \\
    &= \lambda_1 \varphi_1 \varphi_2 + \lambda_2 \varphi_1 \varphi_2 \\
    &= (\lambda_1 + \lambda_2) \varphi_1 \varphi_2.
\end{split}
\end{equation}
This represents that $\varphi_1\varphi_2$ is another eigenfunction with corresponding eigenvalue, $\lambda_1 + \lambda_2$.
The lattice structure implies that, in the numerical implementation, one might obtain multiple eigenpairs, but they might be generated by a smaller set of unique eigenpairs.

% In discrete time setup, similar result can be derived,

% \begin{equation}\begin{split}
%     \koop (\varphi_1(\vx) \varphi_2(\vx)) &= \varphi_1(\vF_t(\vx))\varphi_2(\vF_t(\vx)) \\
%     &= \lambda_1 \lambda_2 \varphi_1(\vx) \varphi_2(\vx)
% \end{split}
% \end{equation}
% where in this case, $\lambda_1 \lambda_2$ is the corresponding eigenvalue.

\paragraph{Koopman bilinear form}
Given a set of eigenpairs $\{(\lambda_i,\varphi_i)\}_{i=1}^n$, the \textit{Koopman Canonical Transform} (KCT) \cite{SURANA2016} of the system of interests Eq. \eqref{eqn:systemEqn_m} is given as 
\begin{equation}
    \dot{\vtvf} = \vtL\vtvf + \sum_{i=1}^p\cL_{\vg i}\vtvf\beta_i\:,
\end{equation}
where $\vtL=\diag([\lambda_1,\cdots,\lambda_n])$, $\vtvf=[\varphi_1,\cdots,\varphi_n]$, and Lie derivatives for parameter dependency term is $\cL_{\vg i}=\vg_i\cdot\nabla$.

Suppose the set of eigenfunctions is sufficiently large, such that $\vtvf$ span an invariant space for $\cL_{\vg i}$, \ie, $\cL_{\vg i}$ can be represented using an $n\times n$ 
matrix $\vD_i$ such that $\cL_{\vg i}\vtvf=\vD_i\vtvf$. Then the KCT can be brought to a bilinear form \cite{Goswami2022},
\begin{equation}\label{eqn:KCT-bi}
    \dot{\vtvf} = \vtL\vtvf + \sum_{i=1}^p\vD_i\vtvf\beta_i\:.
\end{equation}
Often it is difficult to directly obtain the eigenfunctions of $\koop$. Instead, by introducing lifted coordinates via a mapping $\vz=\bar{\vtf}(\vx)$ such that $\vtvf=\vtY^H\vz$, and substituting it into the KCT in bilinear form Eq.~(\ref{eqn:KCT-bi}), we reach the commonly used Koopman Bilinear Form (KBF) \cite{Goswami2022,Jiang2022},
\begin{equation}\label{eqn:kbf}
    \dot{\vz}=\vA\vz+\sum_{i=1}^p\vB_i\vz\beta_i\:,
\end{equation}
%\vtF^H\dot{\vtvf}=
where an eigendecomposition $\vA=\vtG\vtL\vtY^H$ reproduces the Koopman eigenvalues $\vtL$ and $\vB_i=\vtG\vD_i\vtY^H$. The original states are recovered from an inverse mapping $\vx=\tilde{\vtf}^{-1}(\vz)\equiv \tilde{\vty}(\vz)$.

The condition for bilinearization into the form Eq.~(\ref{eqn:kbf}) is restated as follows \cite{Goswami2022}.
Suppose a set of the Koopman eigenfunctions $\{\varphi_1,\varphi_2,\dots,\varphi_n\}$ of the autonomous system forms a invariant subspace of $\cL_{\vg i},i=1,\dots,p$. Then Eq. \eqref{eqn:systemEqn_m} is bilinearizable with an $n$-dimensional state space.

\subsubsection{Extended KBF Model}

In general it might not be possible to write a nonlinear aeroelastic system in an input-affine form of Eq. \eqref{eqn:systemEqn_m}, and one needs to work with a general nonlinear form,
\begin{equation}\label{eqn:org_nlin}
    \dot{\vx} = \vh (\vx; \tilde{\vtb}).
\end{equation}
We propose an extended KBF model to globally linearize Eq. \eqref{eqn:org_nlin}.  To motivate the idea, consider the one-parameter case, i.e., $\tilde{\vtb}=[\tilde{\beta}]$.
Applying Taylor series expansion of Eq. \eqref{eqn:org_nlin} in $\tilde{\beta}$ around a reference value $\beta^*$, one obtains
\begin{align}\nonumber
    \dot{\vx} &= \vh(\vx; \beta^*+\Delta{\beta}) = \vh(\vx; \beta^*) + \frac{\partial \vh}{\partial \beta} \bigg|_{\beta^*}\Delta{\beta} + \frac{\partial^2 \vh}{\partial \beta^2} \bigg|_{\beta^*} \Delta{\beta}^2 + \cdots \\\label{eqn:nlin_taylor}
    &\equiv \vf(\vx) + \vg_1(\vx) \Delta{\beta} + \vg_2(\vx) \Delta{\beta}^2 + \cdots.
\end{align}
Clearly, if one truncate Eq. \eqref{eqn:nlin_taylor} up to order $k$, and define new input parameters $\beta_i=\Delta\beta^i$, $i=1,2,\cdots,k$, then a new KBF model can be formed exactly as in Eq. \eqref{eqn:kbf}.

Extending to multiple input parameters, one can form a general representation as
\begin{equation}\label{eqn:ekbf}
    \dot{\vz}=\vA\vz + \underbrace{\sum_{i_1=1}^p\vB_{i_1}\vz\beta_{i_1}}_{\text{1st-order}} + \underbrace{\sum_{i_1,i_2=1}^p \vB_{i_1, i_2} \vz \beta_{i_1} \beta_{i_2}}_{\text{2nd-order}} + \cdots + \underbrace{\sum_{i_1,\cdots, i_k=1}^p \vB_{i_1, \cdots, i_k} \vz \beta_{i_1}\beta_{i_2} \cdots \beta_{i_k}}_{k\text{th-order}} \equiv \tilde{\vA}(\vtb)\vz,
\end{equation}
which will be hereafter called $k$-th order Extended KBF (EKBF) model. In the case of $k=1$, the model becomes the standard KBF model, and in the case of $p=1$, the EKBF model corresponds to the one-parameter case shown in Eq. \eqref{eqn:nlin_taylor}.
For a fixed choice of the system parameters $\vtb$, the system dynamics can be reformulated as a linear system with system matrix $\tilde{\vA}(\vtb)$.

\subsection{Stability Analysis Using Koopman Theory}\label{ssc_bistab}

An aeroelastic system with a fixed flight condition, i.e., fixed $\vtb$ in our notation, can be represented by an autonomous one.  In this section, we show how exactly the global linear and the nonlinear systems are related to each other, and how the eigenpairs of the global linear system reveal the stability characteristics of the original nonlinear dynamics.  Specifically for aeroelastic applications, one of the most common type of instability is flutter, or Hopf bifurcation.  Hence we will focus on the application of Koopman operator for the stability analysis of two types of dynamical systems: (1) systems with a fixed point (\ie, pre-bifurcation or pre-flutter), and (2) systems with a limit cycle (\ie, post-bifurcation or post-flutter).

\subsubsection{General Concept}\label{sss_gi}
To start with, we review the concept of topological conjugacy between two dynamical systems.  Suppose there are two flow maps, $\vF_t: \bX\mapsto\bX$ and $\tilde{\vF}_t: \bY\mapsto\bY$, and there exists a homeomorphism $\vd: \bX\mapsto\bY$, such that
\begin{equation}
    \vd\circ\vF_t = \tilde{\vF}_t\circ\vd,
\end{equation}
then the two dynamical systems are said to be topologically conjugate to each other.  In the geometrical sense, the conjugacy indicates that, any trajectory generated by $\vF_t$ can be continuously mapped to and from one generated by $\tilde{\vF}_t$; in other words, the dynamics of the two systems are ``equivalent''.

In general, Koopman eigenfunctions provide a way to construct the conjugate map and extend the conjugacy to the attraction basin of the attractor (fixed point or limit cycle) or the end of the interval of existence of the trajectory \cite{Lan2013,Brunton2022KoopmanTheory}.  Conversely, the stability results of the linear system, which is typically easy to obtain, propagate back to the original nonlinear system.  Specifically, the combination of Koopman eigenfunctions and eigenvalues reveals the stability characteristics of the attractor.  If $\varphi_i$ has an eigenvalue whose real part is negative, then the zero level set $M_0=\{\varphi_i(\vx)=0\}$ is asymptotically stable \cite{Mauroy2016}.

In the following the above stability characterization is discussed in detail for the specific cases of fixed point and limit cycle.

\subsubsection{Fixed Point}\label{sss_fp}
Consider the following two dynamical systems:
\begin{compactenum}
    \item A nonlinear system: $\dot{\vx} = \vf(\vx)$ in a domain $\vx\in\bX\subset\bR^r$ with a fixed point at $\vx^*$, that is, $\vf(\vx^*)=0$.  Without loss of generality, let $\vx^*=0$.  The system induces a flow map $\vF_t$, with Koopman eigenpairs $(\lambda_i,\varphi_i)$.
    \item A linear time-invariant system: $\dot{\vy}=\vJ\vy$ in the domain $\vy\in\bY\subset\bR^r$, where $\vJ=\nabla\vf(0)$.  The linear flow map is denoted $e^{\vJ t}$.
\end{compactenum}
The classical Hartman–Grobman theorem \cite{Chicone2006} guarantees the existence of the conjugate map $\vd$ between the two systems, in a sufficiently small neighborhood $N$ of $\vx=0$, when the fixed point is hyperbolic, i.e., when all the real parts of the eigenvalues of $\vJ$ are nonzero.
% Chicone, C. (2006). Ordinary Differential Equations with Applications. Texts in Applied Mathematics. Vol. 34 (2nd ed.). Springer.

Specifically, the existence of conjugate map indicates
\begin{equation}\label{eqn:fpc}
    \vd \circ \vF_t = e^{\vJ t} \vd.
\end{equation}
Consider the case where $\vJ$ is diagonalizable.  Denote eigendecomposition $\vJ=\vW\tilde{\vtL}\vV^H$, where $\vW$ and $\vV$ contain the columns of the right and left eigenvectors and $\tilde{\vtL}$ contains eigenvalues.  Equation \eqref{eqn:fpc} is manipulated further,
\begin{align}\nonumber
    \vd \circ \vF_t &= \vW e^{\tilde{\vtL} t}\vV^H \vd \\\label{eqn:fpd}
    \Rightarrow\quad \vV^H \vd \circ \vF_t &= e^{\tilde{\vtL} t}\vV^H \vd.
\end{align}
Define $\tilde{\varphi}_i=\uv_i^H\vd$, then Eq. \eqref{eqn:fpd} indicates
\begin{equation}
    \tilde{\varphi}_i \circ \vF_t = e^{\tilde{\lambda}_i t} \tilde{\varphi}_i.
\end{equation}
Clearly, $\tilde{\varphi}_i$ is a special Koopman eigenfunction of $\vF_t$ whose eigenvalue $\tilde{\lambda}_i$ coincides with one of the eigenvalues of the linearized system.  Furthermore, since the linear system is $r$-dimensional and hence has $r$ eigenvalues, each eigenvalue would correspond to one Koopman eigenfunction.  In total this correspondence produces $r$ Koopman eigenfunctions, which are referred to as principal eigenfunctions, and the vector of such eigenfunctions is denoted $\vtvf_p$.  From Eq. \eqref{eqn:fpd}, $\vV^H\vd = \vtvf_p$, and the conjugate map can be expressed explicitly as
\begin{equation}
    \vd = \vW\vtvf_p.
\end{equation}

As a side note, $\tilde{\varphi}_i(\vy)=\uv_i^H\vy$ is actually a Koopman eigenfunction of the linear system with eigenvalue $\tilde{\lambda}_i$, since
\begin{equation}\label{eqn:lef}
    \dot{\tilde{\varphi}}_i(\vy) = \uv_i^H\dot{\vy} = \uv_i^H\vJ\vy = \tilde{\lambda}_i \uv_i^H\vy = \tilde{\lambda}_i \tilde{\varphi}_i(\vy).
\end{equation}
Also, $\tilde{\varphi}_i$ and $\varphi_i$ are aligned at $\vx=0$, in the sense that the gradients $\uv_i^H$ and $\nabla\varphi_i(0)$ are collinear, which is one might have expected from the linearization process.

Having established the conjugacy between the nonlinear and linear dynamics, one can leverage the results for the stability analysis.
First, in the fixed point case, i.e., in pre-flutter regime, one can see that the Koopman eigenvalues and eigenfunctions are a natural generalization of the linearized eigenvalues and eigenvectors.
Furthermore, one can leverage the stability criteria in the Koopman context.  Citing the general theorem from Ref. \cite{Mauroy2016}, if there are $r$ Koopman eigenfunctions whose eigenvalues have negative real parts and the vectors $\nabla\varphi_i(0)$ are linearly independent, then the fixed point is asymptotically stable.  When the fixed point is not hyperbolic, i.e., $\vJ$ has at least one eigenvalue whose real part is zero, then one needs to resort to the zero level set approach discussed in Sec. \ref{sss_gi}.

\paragraph{Analytical Example, Part I}
A two-part example on 2D Hopf bifurcation is considered to visualize the behavior of Koopman eigenfunctions.  This example also introduces the concepts of isostable and isochron \cite{mauroy2013isostables}, that are useful for characterizing nonlinear dynamics and its stability.

The dynamics is given by
\begin{subequations}\label{eqn:2d_hopf}
    \begin{align}
        \dot{r} &= r(\mu-r^2)\\
        \dot{\theta} &= \omega(r),
    \end{align}
\end{subequations}
where the frequency $\omega(r)$ can be amplitude dependent in general.  In addition, in Part I we consider the fixed-point case.  The system shows different behaviors depending on the sign of $\mu$. For $\mu \leq 0$, $\dot{r}$ is always negative, which provides a global fixed point attractor at the origin point. Meanwhile, for $\mu > 0$, the origin point becomes an unstable fixed point, and the system forms a limit cycle oscillator at $r=\sqrt{\mu}$, if the system is represented in Cartesian coordinates $(x,y)=(r\cos(\theta),r\sin(\theta))$.

In this section, we consider a simple case, $\omega(r)=2\pi$.  Using a Cartesian representation of Eq.~\eqref{eqn:2d_hopf}, the linearization at the equilibrium point at the origin gives two linearized eigenvalues; as discussed earlier, these linearized eigenvalues correspond to the two principal Koopman eigenvalues $\lambda_{1,1} = \mu +  2\pi i$ and $\bar{\lambda}_{1,1} = \mu - 2\pi i$.
Furthermore, the analytical solutions to the Koopman problem is known \cite{bagheri2013koopman}, and the eigenfunctions are calculated by the nonlinear eigenvalue problem shown in Eq. \eqref{eqn:lie_eqn}
% As mentioned earlier, the principal eigenvalues of the nonlinear system are the ones that correspond to those of the linearized system.
% The linearization of the system around a nominal trajectory $(r_0, \theta_0) = (0, 2\pi t)$ leads to the following linear system,
% \todo{in cartesian}
% \begin{subequations}\label{eqn:lin_org}
%     \begin{align}
%         \delta \dot{r} &= \mu \delta r \\
%         \delta \dot{\theta} &= 0, \quad (\dot{\theta} = 2\pi)
%     \end{align}
% \end{subequations}
% for some perturbation $(\delta r, \delta \theta)$ around the nominal trajectory $(r_0, \theta_0)$. Therefore, one can see that the principal eigenvalues are $\lambda_{1,1} = \mu +  2\pi i$ and its complex-conjugate, $\bar{\lambda}_{1,1} = \mu - 2\pi i$.  Note that the system shows $2\pi$ angular velocity everywhere, so the imaginary part should appear.
\begin{equation}\begin{split}\label{eqn:2d_hopf_eig_form}    \varphi(r, \theta; j, m) = \left(1- \mu r^{-2}\right)^{\frac{-j\alpha_0}{2\mu}} \exp(+i m\theta) \\
    \bar{\varphi}(r, \theta; j, m) = \left(1- \mu r^{-2}\right)^{\frac{-j\alpha_0}{2\mu}} \exp(-i m\theta),
\end{split}
\end{equation}
where $\alpha_0 = \mu$ is the real part of the principal eigenvalues, and $j$ and $m$ are indices that provide the lattice structure; one can come up with new eigenvalues and eigenfunctions using the principal eigenvalues and eigenfunctions as described in Sec. \ref{koopman_lattice}.  The lattice structure of the eigenvalues is formed as $\lambda_{j,m} = j \mu + 2\pi mi$.
% For $j=1$ and $m=1$, the corresponding eigenfunctions are

% \begin{equation}\begin{split}
%     \varphi_{1,1} &\equiv \varphi (r, \theta;1,1) = \left(1- \mu r^{-2}\right)^{-1/2} \exp(+i \theta) \\
%     \bar{\varphi}_{1, 1} &\equiv \bar{\varphi}(r, \theta;1,1) = \left(1- \mu r^{-2}\right)^{-1/2} \exp(-i \theta).
% \end{split}
% \end{equation}

The product of the two eigenfunctions gives
\begin{equation}
    \varphi_{2, 0} = \left(1 - \mu r^{-2} \right)^{-1},
\end{equation}
whose corresponding eigenvalue is $2\mu$, which provides the largest purely decaying mode.  This can also be thought of as the eigenfunction corresponding to $j=2, m=0$. 

% \begin{subequations}\label{eqn:2d_hopf_eigs}
%     \begin{equation}
%     \lambda_1 = \begin{cases}\mu & \mu < 0 \\
%     -2\mu & \mu \geq 0 \;
%     \end{cases}, \quad
%     \varphi_1 = \begin{cases}
%    \left(1-\frac{\mu}{r^2}\right)^{-1/2} & \mu < 0 \\ 
%    \frac{\mu}{r^2} - 1 & \mu \geq 0, r^2 < \mu \\
%    1 - \frac{\mu}{r^2} & \mu \geq 0, r^2 \geq \mu
% \end{cases} 
% \end{equation}
%         \begin{equation}
%     \lambda_{2},\bar{\lambda}_2 = \begin{cases}\mu \pm 2\pi i & \mu < 0 \\
%     -2\mu \pm 2\pi i & \mu \geq 0 \;
%     \end{cases}, \quad
%     \varphi_2, \bar{\varphi}_2 = \begin{cases}
%    \left(1-\frac{\mu}{r^2}\right)^{-1/2}\exp(\pm i\theta) & \mu < 0 \\ 
%    \left(\frac{\mu}{r^2}-1\right) \exp(\pm i \theta) & \mu \geq 0, r^2 < \mu \\
%    \left(1-\frac{\mu}{r^2}\right) \exp(\pm i \theta) & \mu \geq 0, r^2 \geq \mu.
% \end{cases} 
% \end{equation}
% \end{subequations}

% Note that the eigenvalues can also be found by considering the local linearization counterpart of Eq. \eqref{eqn:2d_hopf}. 
%  For $\mu \leq 0$, the system is still in the pre-bifurcation region, and hence the linearization is around the origin.
% \begin{subequations}\label{eqn:lin_org}
%     \begin{align}
%         \delta \dot{r} &= \mu \delta r \\
%         \dot{\theta} &= 2\pi
%     \end{align}
% \end{subequations}

As illustrated in Fig. \ref{2d_eigf_intro_fp}.
The 3D plot in Fig. \ref{2d_eigf_intro_fp} (a) shows the absolute value of the eigenfunction, $\mathcal{V} = |\varphi_{1,1}|$, and is called isostable; the 2D contour indicates $|\nabla \mathcal{V}|$, where the center has stronger gradient.  Meanwhile, the phase of the eigenfunction, i.e., $\arg{(\varphi_{1,1})}$ is called isochron.  The role of isostable is discussed below, while the role of isochron is explained later in Part II of this example.

The isostable provides the useful information on the asymptotic convergence of the system at the global level, which is illustrated in Fig. \ref{2d_eigf_intro_fp}.
A set of points that shares the same level of isostable have the same asymptotic convergence toward the stable fixed point, and the stable fixed point corresponds to the point of zero level set.  The isostable is in fact a Lyapunov function that describes the global stability of the system \cite{Mauroy2016}. It can also be treated as non-negative observable, and hence under the action of the Koopman operator, the value decreases in time, i.e., $\koop \mathcal{V}(\vx) < 0$ for all $ \vx \neq \vx^*$. This is because by definition of $\mathcal{V}(\vx)$ and Eq. \eqref{eqn:koopman-eigen}, 
\begin{equation}
    \frac{d}{dt} \mathcal{V}(\vx) = \sigma \mathcal{V}(\vx) < 0,
\end{equation}
where $\sigma = Re(\lambda) < 0.$ for a stable system. In this setting, the evolution of trajectory under nonlinear map $\vf$ can be understood by the evolution of this Lyapunov function as shown in Fig. \ref{2d_eigf_intro_fp} (b). The Lyapunov function essentially evolves in time by
\begin{equation}
    \dot{\mathcal{V}}(\vx(t)) = \nabla \mathcal{V}(\vx) \cdot \vf(\vx),
\end{equation}
which quantifies the similarity between the gradient vector of Lyapunov function and the vector field of the nonlinear system. In the case of Fig. \ref{2d_eigf_intro_fp} (b), the angle between the two vectors becomes greater as it gets close to the center fixed point. This contributes to stronger attraction towards the center, although the quantity also depends on the magnitude of the vectors.

\begin{figure}[H]
    \centering
    \includegraphics[width=\textwidth]{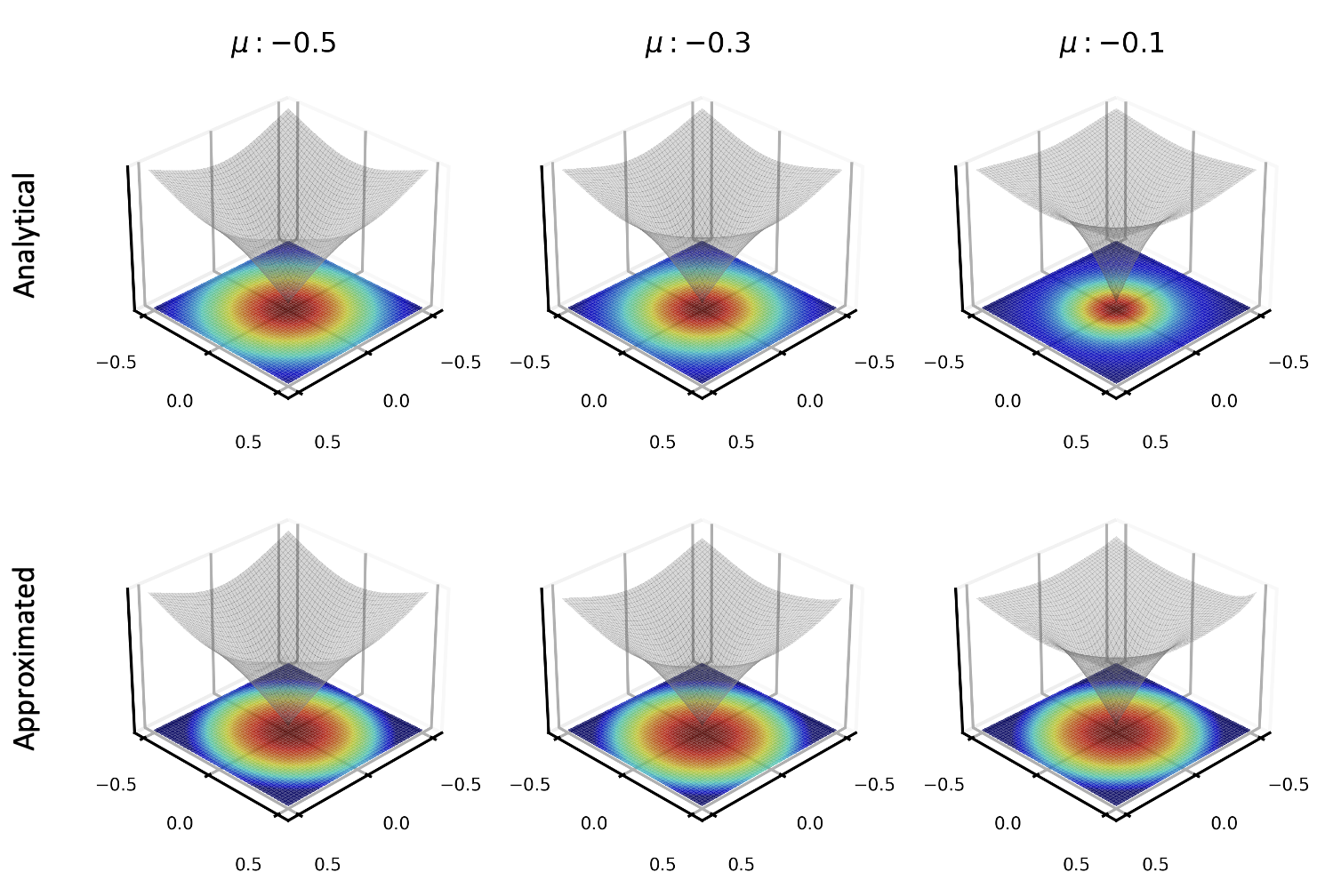}
    \caption{Isostable for different system parameter $\mu$ in the pre-bifurcation regime.}\label{2d_eigfs_prebif}
\end{figure}

% \todo{Update this}
In Fig. \ref{2d_eigfs_prebif}, both isostable and its magnitude of the gradient is shown in a similar manner.  The approximated eigenfunctions are calculated by standard EDMD algorithm \cite{williams2015data} using 5-th order monomial lifting.  The color scale is individually adjusted for each case and is not shared across the different cases.  When $\mu=-0.5$ the slope of eigenfunction is relatively more gentle than other eigenfunctions at different $\mu$. As $\mu$ increases, the slope gets steeper near the center, and the points that are further away from the center gets flattened.  The numerical approximation seems to capture the isostable while there is non-negligible error in the gradient especially near the bifurcation point.

% Moreover, the isochron is also calculated and colored on the bottom surface of the plot. In this case, the isochron extends radially outwards. This is simply due to the fact that the angular frequency is state-independent ($\dot{\theta} = 2\pi$). These properties of eigenfunction are successfully captured by the numerical algorithm.

% In addition, the phase of the eigenfunction, i.e., $\arg{(\varphi)}$ is represented as a color map on the bottom surface of the function in Fig. \ref{2d_eigf_intro} (a).
% A set of points that has the same color has same phase in the oscillation. This set is called isochron in literature. 

\begin{figure}
\centering
\insertfigs{pics/pre_bifurcation}{0.43}{Globally stable fixed point and the isostable as Lyapunov function.}
\insertfigs{pics/pre_bifurcation_eigf_dyn}{0.48}{The contours of gradient of isostable.}
\caption{A 2D Hopf bifurcation example in the pre-bifurcation regime with stable fixed-point.}\label{2d_eigf_intro_fp}
\end{figure}

\subsubsection{Limit Cycle}
Next, the analysis for the post-bifurcation regime is presented.

Formally, consider a nonlinear system, $\dot{\vx}=\vf(\vx)$ on a domain $\vx\in\bX\subset\bR^r$, $\vx\in\bX\subset\bR^r$, that admits an orbit of period $T$, i.e., the limit cycle $\vx^*(t)=\vx^*(t+T)$.  While a detailed Koopman analysis is provided in Ref. \cite{Mauroy2016} based on frequency analysis, here we provide a simplified discussion using the Poincar\'{e} map and the time-delayed coordinates.

The Poincar\'{e} map is a well-known tool for the study of limit cycle systems.  Formally, denote the map $\vp:\bP\mapsto\bP$, where $\bP\subset\bX$ is a hyperplane that intersect with the limit cycle at a certain time $t_p$; denote the intersecting point as $\vx_p$.  When a trajectory of the nonlinear dynamics passes through a point $\vx_n\in\bP$, $\vp$ maps $\vx_n$ to the next point $\vx_{n+1}\in\bP$ where the trajectory intersects with $\bP$ again.  Clearly, if the limit cycle is stable, $\lim_{n\rightarrow\infty}\vp^n(\vx_0)=\vx_p$, as long as $\vx_0$ is in the attraction basin of the limit cycle.

Following Poincar\'{e} map, define its linearization at $\vx=\vx_p$, so that $\vy_{n+1}=\vP\vy_n$, where $\vP=\ppf{\vp}{\vx}$.  Following a similar argument in the previous section, the linearized system in $\vy$ is topologically conjugate to the original nonlinear Poincar\'{e} map.  Further, with the help of Koopman theory, the equivalence is extended to the entire attraction basin of the limit cycle \cite{Mauroy2016}.  Again, the conjugacy map $\vy=\vd(\vx)$ between the two systems is achieved using the Koopman eigenfunctions of the Poincar\'{e} map.

For the case of limit cycles, the Koopman eigenvalues correspond to the Floquet exponents of the original system, that determine whether and how the trajectory converges to the limit cycle.  Specifically, if $\vP\vu_0=\lambda\vu_0$, then a unit eigenvector evolves under $\vP$ as
\begin{equation}
    \vu(t) = \exp(\mu t)\vu_0,
\end{equation}
where $\mu=\ln(\lambda)/T$ is the Floquet exponent for $\vu_0$.

Subsequently, consider a time-delay embedding of coordinates, sampled $N$ times over the period of limit cycle uniformly with time step size $\Dt=T/N$.  Using the conjugacy mapping, the dynamics near the limit cycle can be represented as a linear system,

\begin{equation}
    \begin{bmatrix}
        \vy(t+\Dt) \\
        \vy(t+2\Dt) \\
        \vy(t+3\Dt) \\
        \vdots \\
        \vy(t+N\Dt)
    \end{bmatrix} =
    \begin{bmatrix}
        \vO & \vI & \vO & \hdots & \vO \\
        \vO & \vO & \vI & \hdots & \vO \\
        \vO & \vO & \vO & \ddots & \vO \\
        \vdots & \vdots & \vdots & \ddots & \vI \\
        \vP & \vO & \vO & \hdots & \vO \\
    \end{bmatrix}
    \begin{bmatrix}
        \vy(t) \\
        \vy(t+\Dt) \\
        \vy(t+2\Dt) \\
        \vdots \\
        \vy(t+(N-1)\Dt)
    \end{bmatrix},
\end{equation}
or in short,
\begin{equation}
    \cY_N = \cP\cY_{N-1}.
\end{equation}
The eigenpairs of $\vP$ and $\cP$ are closely related.  For an eigenpair of $(\lambda,\vu)$ of $\vP$,
\begin{equation}
    \cU^\top=[\vu(0)^\top, \vu(k\Dt)^\top, \cdots, \vu(k(N-1)\Dt)^\top],
\end{equation}
is an eigenvector of $\cP$ with eigenvalue $\exp(k\mu\Dt)$.  One may check the expression by definition.  Finally, in practice, the time delay embedding just needs to be sufficiently long to encompass the period of the limit cycle to capture the dynamics.

Having established the relation between the nonlinear limit cycle system and a linear system, the validity of which holds in the attraction basin of the limit cycle, one can leverage the relation for the stability analysis, in a manner similar to the previous section.  For a $r$-dimensional system, if there are $r-1$ Koopman eigenfunctions whose eigenvalues have negative real parts and the gradients of the eigenfunctions are linearly independent, then the limit cycle is asymptotically stable.

\paragraph{Analytical Example, Part II}

Continuing the earlier 2D example, this part considers the post-bifurcation regime with a stable limit cycle.

In this case, the principal eigenvalues are obtained by linearizing the system around the limit cycle amplitude, $r=\sqrt{\mu}$,
\begin{subequations}\label{eqn:lin_lco}
    \begin{align}
        \delta \dot{r} &= - 2\mu \delta r \\
        \delta \dot{\theta} &= 0,
    \end{align}
\end{subequations}
where the linearized eigenvalues are $-2\mu$ and $0$, and the former correspond to the Floquet exponent.

For the principal Koopman eigenvalue $\lambda_{1} = -2\mu$, the associated eigenfunction is
% Note that for a 2-dimensional system ($r=2$), there is one Koopman principal eigenfunction and eigenvalue pair ($r-1=1$) where the principal eigenvalue is $\lambda_{1} = -2\mu $.  Therefore, $\alpha_0=-2\mu$ in Eq.~\eqref{eqn:2d_hopf_eig_form}.  On the Poincar\'{e} map, the system only sees the radial dynamics, $\dot{r} = r(\mu - r^2)$.
% Then the principal eigenfunction whose eigenvalue corresponds to the Floquet exponent is found as follows 
\begin{equation}
    \varphi_{1} = 1- \mu r^{-2},
\end{equation}
which is the radial component of the Eq.~\eqref{eqn:2d_hopf_eig_form} with $\alpha_0=-2\mu$.  This eigenfunction only contains the decaying component around the fixed point on the Poincar\'{e} map.  Note that there is a pair of eigenfunctions that provide the limit cycle oscillations, $\varphi_2 = \exp{(+i\theta)}$, and $\bar{\varphi}_2 = \exp{(-i\theta)}$.  

Figure \ref{pics/post_bifurcation} shows the isostable, $|\varphi_1|$, as a 3D surface and the isochron, $\arg(\varphi_2)$, as a 2D contour.  This time zero level set $M_0 = \{\varphi_i(\vx) = 0\}$ consists of limit cycle. The meaning of isostable remains the same as in the fixed-point case, which governs the convergence rate to the attractor. However, the isochron provides more valuable insights into the asymptotic convergence of limit cycle system. In Fig. \ref{pics/isocrhon_demo}, trajectories originating from the blue initial points evolve in a counterclockwise direction and gradually converge toward the limit cycle. Points along the trajectory are marked at regular time intervals, with identical markers indicating positions reached after the same elapsed time from the initial points. One can see that the same markers are on the same isochron meaning that they are at the same phase of oscillation even in the presence of attractor.  As a result, the points of the same isochron value eventually merge onto the same point on the limit cycle as time approaches infinity. This phase information of limit cycle dynamics endowed by isochron is consistent, if it exists, no matter how complex the dynamical system is, which makes it useful to extract the phase information of high-dimensional and complex dynamical systems \cite{taira2018phase}.

\begin{figure}[H]
    \centering
    \includegraphics[width=\textwidth]{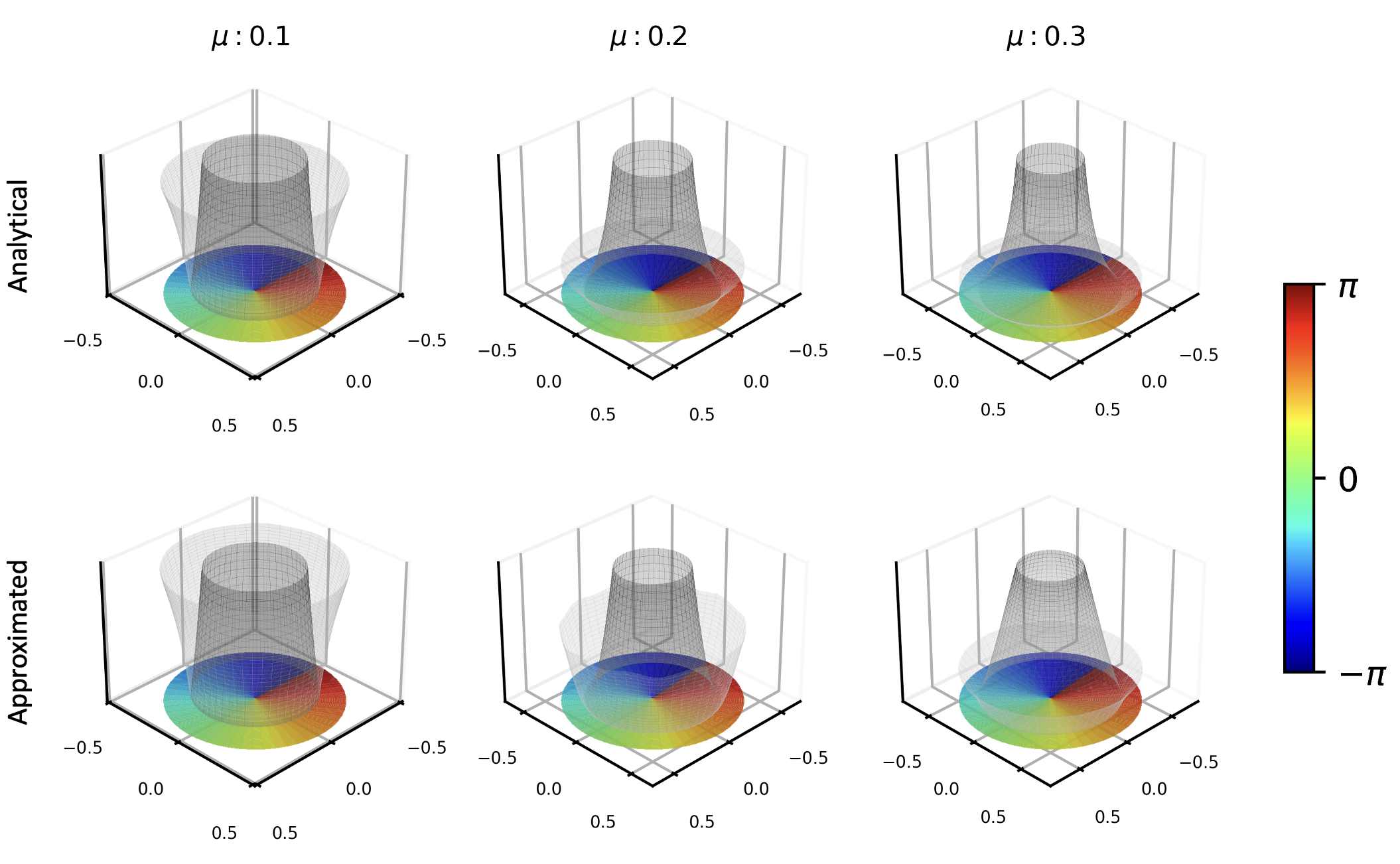}
    \caption{Isocrhon for different system parameter $\mu$ in the post-bifurcation regime.}\label{2d_eigfs_postbif}
\end{figure}

% \todo{Possible relocation destination}
In Fig. \ref{2d_eigfs_postbif}, the analytical and approximated eigenfunctions are compared.   A set of points at which the value of eigenfunction is zero forms the limit cycle, and any neighboring points are attracted by the limit cycle according to the eigenfunction dynamics, $\dot{\varphi} = \lambda \varphi$. Furthermore, similar to the previous case, the eigenfunction forms $360^{\circ}$ phase in the $x$-$y$ state space. The approximated eigenfunctions capture the property of the analytical eigenfunctions such as the size and shape of limit cycle.

\begin{figure}
\centering
\insertfigs{pics/post_bifurcation}{0.43}{Globally stable limit cycle and the contours of isochron.}
\insertfigs{pics/isocrhon_demo}{0.48}{Isochron to indicate phase alignment.}
\caption{A 2D Hopf bifurcation example in the post-bifurcation regime with stable limit cycle.}\label{2d_eigf_intro_lc}
\end{figure}

\subsection{Flutter Prediction Algorithm via EKBF Model}\label{sec:num_alg}

Through Secs. \ref{ssc_dat} and \ref{ssc_bistab}, the formulation of the EKBF model and the connection of EKBF to local stability are described.  The last piece of the Koopman framework is the numerical implementation of the EKBF model and the EKBF-based flutter prediction.
First, the three key components of the algorithm are presented, including (1) the typical choices of nonlinear coordinate transformation between physical and Koopman states, (2) the typical approach for estimating the system matrices in EKBF from data, and (3) two spectral filtering methods to mitigate the adverse effects of data noise.  Finally, the components are assembled to produce the complete flutter prediction algorithm.

\subsubsection{Choice of Coordinate Transformation}

To identify a finite-dimensional Koopman operator from the data, the first step is to define the mapping $\tilde{\vtf}$ to transform the states $\vx$ to $\vz$ and its inverse mapping $\tilde{\vty}$.  Three typical approaches include:

\begin{compactenum}
    \item Use explicit nonlinear features, e.g., a set of polynomials, as has been done in standard EDMD-type methods.  An example choice of features are
\begin{equation}
    \tilde{\vtf}(\vx) = [1, \vx, \vx^{\otimes 2}, \vx^{\otimes 3}, \cdots],
\end{equation}
    where $\Box^{\otimes n}$ denotes all $n$th order monomials.  For this example, the inverse $\tilde{\vty}$ is just a linear mapping that extracts out the first $r$ element of $\vz$.
    \item Use a time-delay embedding (TDE) of $\vx$, i.e., \cite{Svenkeson2016,Glaz2017}
\begin{equation}
    \tilde{\vtf}(\vx) = [\vx, \vx(t-\Delta t), \vx(t-2\Delta t), \cdots],
\end{equation}
    in which case the same inverse as the previous case can be used.  The TDE is typically useful when the known physical states $\vx$ are an incomplete representation of the system dynamics, e.g., in experimental measurements, and TDE helps recovering the missing dynamics.
    \item Use an autoencoder architecture, where the mapping and its inverse are both neural networks, denoted $\vz=\tilde{\vtf}(\vx;\vtq)$ and $\vx=\tilde{\vty}(\vz;\vtq)$, where $\vtq$ are learnable parameters, and $\tilde{\vtf}$ and $\tilde{\vty}$ can be simply fully-connected neural networks.
\end{compactenum}

\begin{figure}[H]
    \centering
    \includegraphics[width=0.5\textwidth]{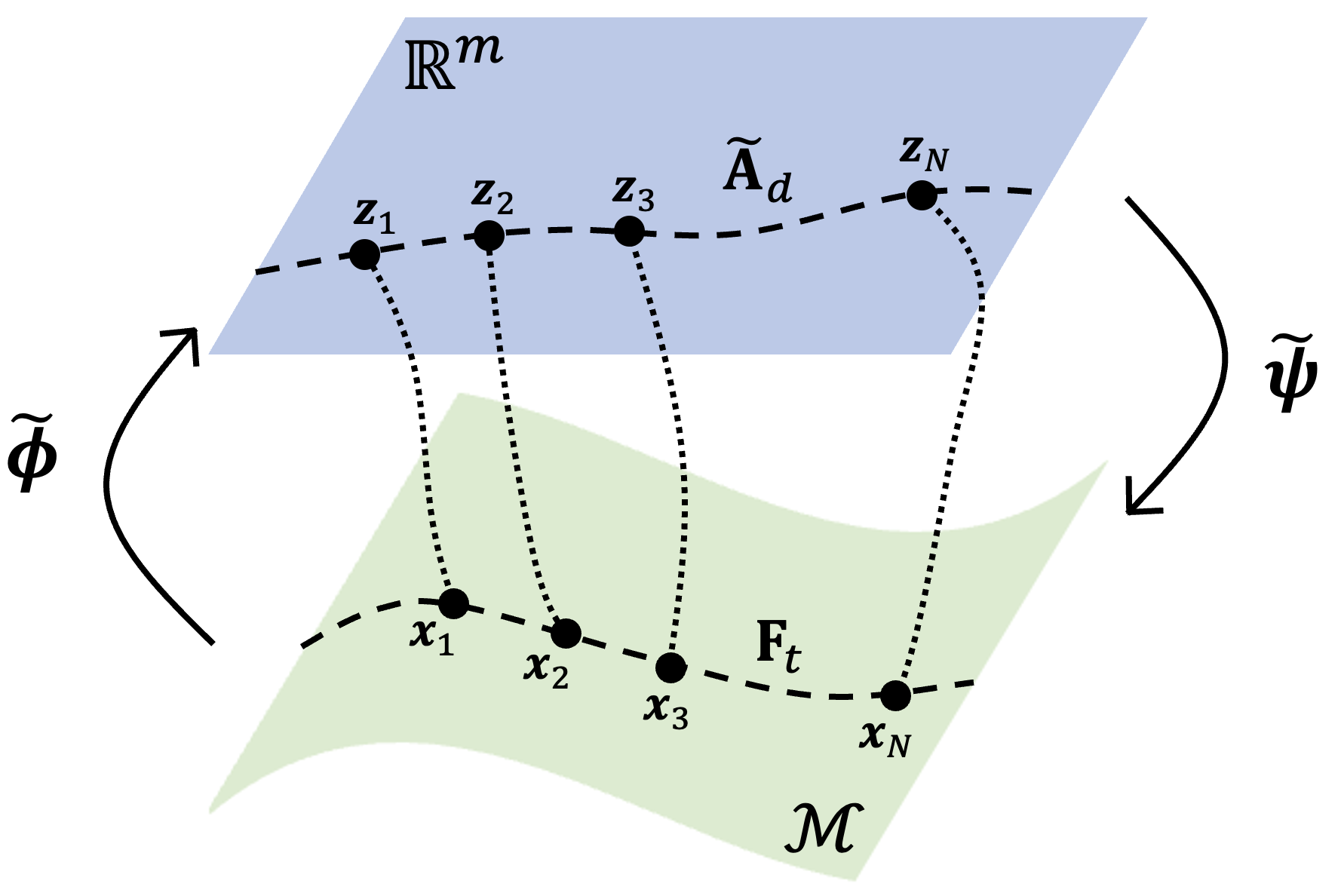}
    \caption{Figure describing the finite dimensional Koopman operator for nonlinear dynamical systems.}\label{KoopmanLifting}
\end{figure}

In Fig. \ref{KoopmanLifting}, the schematics that shows how $\tilde{\vtf}$ and $\tilde{\vty}$ work is provided. Suppose there are discrete samples of trajectory $\vx(t)$, denoted as $\{\vx_1, \vx_2, \dots, \vx_N\}$. Then by mapping through $\tilde{\vtf}$, one can obtain corresponding discrete samples of $\vz(t)$, denoted as $\{\vz_1, \vz_2, \dots, \vz_N \}$. One can apply 
$\tilde{\vty}$ to recover the original state, $\vx$.

Note that, regardless of the choice of coordinate transformation, one necessity condition for the choice of nonlinear mapping is that the linear space of $\nabla_{\vx}\tilde{\vtf}$ should be a subset of that of $\tilde{\vtf}$, so that the Eq. \eqref{eqn:KCT-bi} is well defined.

\subsubsection{Learning the System Dynamics}
Once the nonlinear mapping is chosen, one can employ standard least squares techniques to fit the $\vA$, $\vB_{i_1}$ $\cdots \vB_{i_1, \cdots, i_k}$,  matrices from data \cite{Goswami2022}.  For conciseness and without loss of generality, the one-parameter case is discussed next.

A typical approach to learning the EKBF starts with the time discretization of Eq. \eqref{eqn:ekbf} given a step size of $\Delta t$. Consider a discrete-time $k$-th order EKBF model,
\begin{align}\nonumber
    \vz_{n+1} &\approx (\vI + \tilde{\vA}(\beta)\Delta t) \equiv \tilde{\vA}_d(\beta)\vz_n  \\\label{eqn:disc_ekbf}
    &= \left(\vA_d + \vB_{d, 1}\beta + \cdots + \vB_{d, 1^{(k)}} \beta^k\right) \vz_n,
\end{align}
where $\vA_d$ and $\vB_{d,1}, \cdots, \vB_{d, 1^{(k)}}$ are discrete-time counterparts of $\vA$, $\vB_{1}, \cdots, \vB_{1^{(k)}}$ respectively, and the superscript $(k)$ indicates $k$ times repetitive indices. Then, one can rewrite Eq. \eqref{eqn:disc_ekbf} as a matrix form,

\begin{equation}\label{eqn:lin}
    \vz_{n+1} = [\vA_d, \vB_{d, 1}, \cdots, \vB_{d,1^{(k)}}]
    \begin{bmatrix}
        \vz_n \\ \vz_n\beta \\ \vdots \\ \vz_n\beta^k
    \end{bmatrix} \equiv \vtG \vtx_n.
\end{equation}

For a total of $N$ steps, one may form a linear equation,
\begin{equation}
    [\vz_{N}, \vz_{N-1}, \cdots, \vz_2] = \vtG [\vtx_{N-1}, \vtz_{N-2}, \cdots, \vtz_1],\quad \text{ or }\quad \vZ=\vtG\vtX,
\end{equation}
where $\vtG = [\vA_d, \vB_{d, 1}, \cdots, \vB_{d,1^{(k)}}]$. Clearly, one can obtain the EKBF system matrices via least squares, $\vtG=\vZ\vtX^+$, where $\Box^+$ denotes pseudo-inverse. Then, the sequential block matrices in $\vtG$ corresponds to $\vA_d, \vB_d, \cdots, \vB_{d, 1^{(k)}}$. Lastly, one can recover the continuous-time matrices from the discrete-time counterparts.  For multiple system trajectories sampled from different initial conditions, one can adopt the ensemble approach introduced in Ref. \cite{tu2013dynamic}.

\subsubsection{Spectral Filtering}

For flutter analysis and prediction, it is critical to obtain accurate spectral properties, including the eigenvalues and eigenfunctions, to characterize the stability.  However, the data-driven identification of Koopman models often result in spurious modes, due to the numerical byproduct of least squares as well as the noise in data \cite{colbrook2023residual}.
Therefore, it is necessary to identify the most plausible modes by applying certain filtering technique.
Next, two spectral filtering techniques are described.  One is the so-called Residual Dynamic Mode Decomposition (ResDMD) algorithm \cite{colbrook2023residual}, that is applicable to diagnostic cases, where one only needs to extract flutter information from existing data at a given operating condition.  The other is motivated by the modal assurance criterion (MAC) from classical flutter analysis, and is applicable to predictive cases, where the EKBF needs to extrapolate to new unknown flutter parameters.

\paragraph{ResDMD algorithm}
The core idea of ResDMD is to select the eigenpairs $(\lambda,\varphi)$ that most accurately satisfy the eigenvalue problem of the Koopman operator, i.e., $\koop \varphi = \lambda \varphi$ by computing the spectral residue of the Koopman operator. Specifically, given the data matrices, $\vZ_1= [\vz_{N-1}, \vz_{N-2}, \cdots, \vz_1]$ and $\vZ$, and an eigenvalue and left-eigenvector pair, i.e, $(\lambda, \vty)$ of $\bar{\vA}(\beta),$ the squared residue is a measure of following quantity,
\begin{equation}\label{eqn:resdmd}
    res(\lambda, \vty)^2 = \frac{\vty^*[\vZ \vW \vZ^\top -\lambda (\vZ_1 \vW \vZ^\top)^\top -\bar{\lambda} (\vZ_1 \vW \vZ^\top) + |\lambda|^2 \vZ_1 \vW \vZ_1^\top]\vty}{\vty^* (\vZ_1 \vW \vZ_1^\top) \vty},
\end{equation}
where $\vW \in \mathbb{R}^{(N-1)\times(N-1)}$ is a diagonal quadrature weight. One can set $\vW = \vI$ for a regular equally weighted integration. % The bar over the symbol, $ \bar{\Box} $ stands for complex-conjugate.
Once the residue is computed for each eigenvalue, a filtering criterion can be applied. Specifically, a threshold, $\epsilon$ can be defined such that modes with $res(\lambda, \vty) > \epsilon$ are discarded. Alternatively, a fixed number of modes can be retained by selecting those with the lowest residue values.
Readers interested in further details are referred to Refs. \cite{colbrook2023residual, colbrook2024rigorous}.

\paragraph{Dual MAC filtering}
The ResDMD filtering is not applicable when data are unavailable, especially if one needs to extrapolate the flutter parameter of an EKBF model to predict the flutter boundary. To address this, one can use modal assurance criterion (MAC) \cite{pastor2012modal, desforges1996mode}. In the extrapolation regime, one can track the filtered modes in the training regime by measuring the similarity between the modes.

Specifically, suppose at parameter $\beta_1$ a mode $\uv_1$ has been selected by filtering, e.g., via ResDMD, and at a neighboring parameter $\beta_2$ there is a candidate mode $\uv_2$.
The MAC of $\uv_1$ and $\uv_2$ is calculated as
\begin{equation}\label{eqn:mac_def}
    \text{MAC}(\mathbf{v}_1, \mathbf{v}_2) = \frac{|\mathbf{v}_1^* \mathbf{v}_2|^2}{\norm{\mathbf{v}_1}_2 \norm{\mathbf{v}_2}_2},
\end{equation}
where $\norm{\cdot}_2$ is $L^2$ norm.  When MAC value is close to 1, the two vectors $\mathbf{v}_1$ and $\mathbf{v}_2$ are close to be collinear, which indicates $\uv_2$ is likely a plausible mode continued from $\uv_1$; otherwise, when MAC value is closer to 0, then the two vectors are nearly perpendicular to each other and $\uv_2$ is either irrelevant to $\uv_1$ or a spurious mode.

Suppose that the system has a dependency on a single parameter, i.e., $\beta$, and there is a sequence of parameters, $\beta_1, \beta_2, \cdots, \beta_{p}$. In the context of flutter dynamics, they can be a sequence of flight speeds. Then one can identify a series of eigendecompositions $\tilde{\vA}(\beta) = \vtU(\beta) \vtL(\beta) \vtY^*(\beta)$ in Eq. \eqref{eqn:ekbf}. Let $\vtU_i \equiv \vtU(\beta_i)$ and assume that the eigenvectors are normalized. Then, one can calculate mode-wise MAC,
\begin{equation}\label{eqn:mac_md}
    \text{MAC}(\vtU_i, \vtU_{i+1}) = |\vtU_i^* \vtU_{i+1}|^2 ,
\end{equation}
where the denominator in Eq. \eqref{eqn:mac_def} is ignored since the normalized eigenvectors are considered, and $|\cdot|$ indicates element-wise absolute value as the product of two matrices may be still complex-valued.  Then, given the parameter sequence, mode tracking can be achieved by sequentially calculating MAC and neglect the modes whose MAC value is less than a certain threshold.  Lastly, one can apply the same idea for the left-eigenvectors, $\vtY$ if needed.

It is found that considering both the right and left eigenvectors in the MAC calculation provides a more robust result.  This motivates the dual MAC algorithm, which considers both eigenvectors simultaneously.  In this case, only the modes that exceed the MAC threshold value for both right and left eigenvectors are retained. For example, let $1-\epsilon$ be a threshold of MAC algorithm and denote $\vtu_i^j, \vty_i^j$ as $j$-th column of $\vtU_i$ and $\vtY_i$ respectively. For any mode indices, $q$ and $r$, the $r$-th mode at $\beta_{i+1}$ is retained only when $\text{MAC}(\vtu_{i}^q, \vtu_{i+1}^r) > 1- \epsilon$, and $\text{MAC}(\vty_{i}^q, \vty_{i+1}^r) > 1- \epsilon$.
% However, if $\text{MAC}(\vtu_{i}^q, \vtu_{i+1}^r) > 1- \epsilon$, but $\text{MAC}(\vty_{i}^q, \vty_{i+1}^r) \leq 1- \epsilon$, then the $r$-th mode at $\beta_{i+1}$ is neglected.

\subsubsection{Flutter Prediction Procedure}

\begin{figure}[H]
    \centering
    \includegraphics[width=\textwidth]{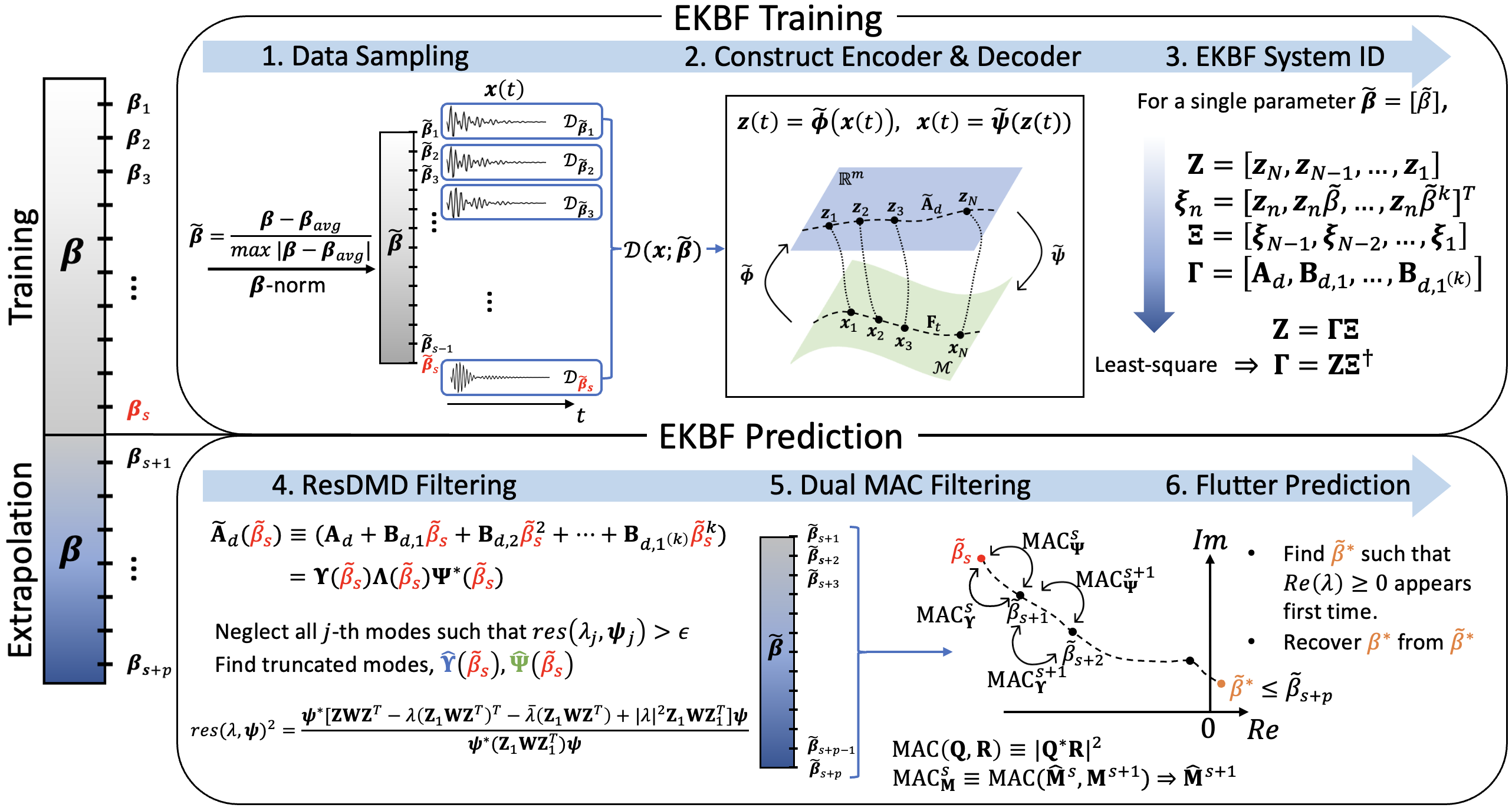}
    \caption{Overview of EKBF flutter prediction algorithm.}\label{ekbf_fp}
\end{figure}

Finally, we present the flutter prediction procedure via EKBF model, as illustrated in Fig.~\ref{ekbf_fp}. In this paper, only the scalar parameter is considered. However, one can still apply the same procedure for multi-parameter case as well. Consider a dataset of trajectories, $\cD = \{\cD_{\beta_1}, \cD_{\beta_2}, \hdots, \cD_{\beta_s}\}$ where $\cD_{\beta_i}$ stands for the dataset generated by a particular parameter $\beta_i$, and assume $\beta_1 < \beta_2 < \cdots < \beta_s$.
Also, suppose one would like to perform flutter prediction for $\beta>\beta_s$, e.g., at parameters $\beta_{s+1}<\beta_{s+2}<\cdots<\beta_{s+p}$.

The algorithm for flutter prediction based on EKBF is formalized as follows:
\begin{compactenum}
    \item Normalize the trajectory data as well as the parameters for training, which is necessary to achieve better numerical behavior.
    \item Define a coordinate transformation to construct $\vz$ in Eq. \eqref{eqn:ekbf}. Typical choice of encoder is monomial lifting, delay embedding, or both.
    \item Determine the order of the EKBF model, $k$.  Typically higher order provides more accurate result but can be computationally inefficient.  Then, find system matrices $[\vA_d, \vB_{d, 1}, \cdots, \vB_{d,1^{(k)}}]$ in Eq. \eqref{eqn:disc_ekbf}, and convert to their continuous-time counterparts.
    \item Apply ResDMD algorithm using Eq. \eqref{eqn:resdmd} to eliminate the spurious modes and obtain the principal eigenvalues. 
    \item Starting from $\beta_{s+1}$, apply MAC defined in Eq. \eqref{eqn:mac_md} to keep track of the same eigenvalue branch found in the previous step.  We specifically use dual MAC for robustness.
    \item Find $\beta^*$ such that the real part of the eigenvalue of the EKBF model at $\beta^*$ crosses over the imaginary axis, indicating the flutter boundary.
\end{compactenum}
% \todo[linecolor=red,backgroundcolor=red!25,bordercolor=red]{EKBF Python package release? (Not ready yet though)}
% \todo{Don't have to be right now.}

\section{Numerical Results}\label{sec2}

In this section, we first revisit the 2D academic example to show how the flutter analysis is performed based on the EKBF model. In the next problem, we will demonstrate the effectiveness of the EKBF model under more realistic circumstance of a panel flutter problem.

\subsection{A 2D Academic Example}

In this section,  we consider a scenario that is more challenging than one presented in Sec. \ref{ssc_bistab} in two aspects.
One challenge is the increased nonlinearity by choosing $\omega(r) = 2\pi/(1+r^2)$, so that the phase is affected by its radial position from the origin. Therefore, the isochron does not radially extend outwards, but there is slight variation as a form of given $\omega$. The linearized system is described as follows
\begin{subequations}\label{eqn:lin_lco_case2}
    \begin{equation}
        \delta \dot{r} = - 2\mu \delta r\label{eqn:lin_lco_case2a}
    \end{equation}
    \begin{equation}
        \dot{\theta} = \frac{2\pi}{1+\mu}\label{eqn:lin_lco_case2b}.
    \end{equation}
\end{subequations}
The other aspect of challenge is the partial measurement.
In many cases, one does not have access to all the states as measurement, e.g., in experiments and during flight. In this scenario, constructing approximate Koopman operator using monomial lifting has been found to be difficult to restore the missing dynamics.

Given partial measurement, the dynamics can be recovered by incorporating delayed coordinates in the state. Suppose following discretized steps, $\{r_k, r_{k+1}, \cdots, r_{k+d}, r_{k+d+1}\}$ where through $d+1$ time steps, the system returns to the original Poincaré map (first returning map). Then this map of the system, $p: \bP\mapsto\bP$ can be obtained by solving Eq. \eqref{eqn:lin_lco_case2a}. The derived Poincaré map is as follows
\begin{equation}\label{eqn:ex1_poinrcare}
    r_{k+d+1} = p(r_k) :=  \left(\frac{1}{\mu} + \left(\frac{1}{r_k^2} - \frac{1}{\mu}\right)\exp{(-2\mu T)}\right)^{-1/2},
\end{equation}
where $T = T(r)$ is the period of the map and can be found by solving Eq. \eqref{eqn:lin_lco_case2b}.  Consider a simple scenario where $T$ is constant; this would have been true if $\dot{\theta}$ dynamics were constant, e.g., $\dot{\theta}=2\pi$. One can expand the expression to obtain KBF model or EKBF model by truncating at a certain order of the parameter. For example, The 2nd order EKBF with 2nd order lifting model will be a form of

\begin{equation}
    r_{k+d+1} = c_{0, 0} + c_{1, 0} r_k + c_{1, 1} \mu r_k + c_{2, 0} r_k^2 + c_{2, 1} r_k^2 \mu + c_{2, 2} r_k^2 \mu^2,
\end{equation}
where $c_{i, j}$ can be found by Taylor expansion of the Poincaré map shown in Eq. \eqref{eqn:ex1_poinrcare} in $r_k$ and $\mu$. If $T$ were not constant, then this would be again approximated by a set of delayed states embedded in the observable. The higher-order lifting can be especially useful when the state is measured in other coordinate system such as Cartesian coordinate so that the model needs to capture the non-linearity introduced by the transformation. Note that to account for $c_{0,0}$, one needs to have a constant term in the observable of the encoder.

Using sample trajectories at different values of $\mu$, the Koopman eigenvalues are estimated using the methods in Section \ref{ssc_dat}.  The states are in the Cartesian coordinate, $\vx = [x, y]$ instead of polar coordinate, and the $x$ component is retained as the partial measurement.
 To account for the additional non-linearity introduced in the transformation, we use 3rd-order lifting in the observable. The data are sampled such that $\Delta t = 0.1$ sec. The number of delay embedding is 50, and 4th-order EKBF model (i.e., up to $\mu^4$) is used for training.
The range of the parameters considered in this experiment is $\mu \in [-0.5, 0.5]$ with $\Delta \mu = 0.05$. Once the model is trained, the eigenvalues are extrapolated for $\mu = [0.55, 0.60, 0.65, 0.70]$.

\begin{figure}
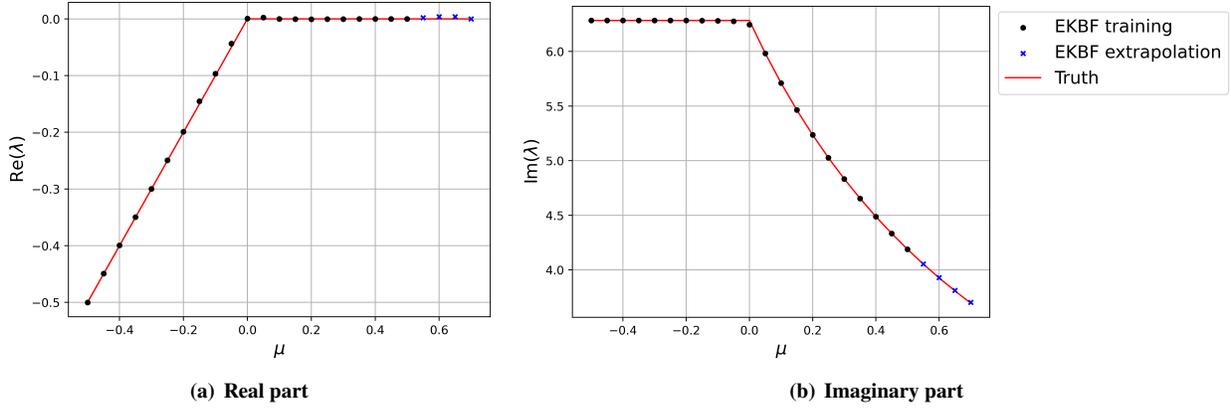

\centering
\insertfigs{pics/re_lambda_1_ext}{0.4}{Real part}
\insertfigs{pics/im_lambda_1_ext}{0.58}{Imaginary part}
\caption{Real and imaginary part of the oscillatory mode, $\lambda_1$ for $-0.5 \leq \mu \leq 0.5$ (training) and $0.55 \leq \mu \leq 0.7$ (extrapolation).}
\label{re_im_eig1}
\end{figure}

\begin{figure}
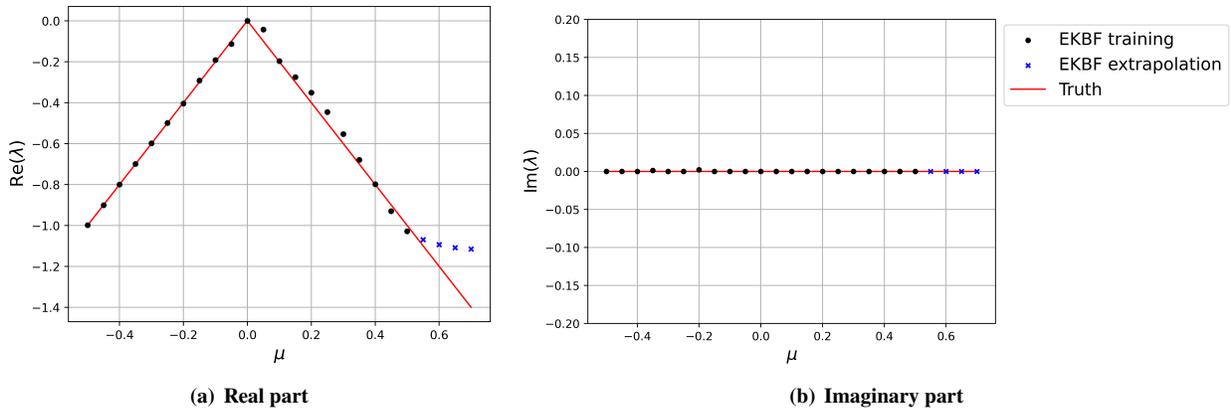

\centering
\insertfigs{pics/re_lambda_2_ext}{0.4}{Real part}
\insertfigs{pics/im_lambda_2_ext}{0.58}{Imaginary part}
\caption{Real and imaginary part of the purely decaying mode, $\lambda_2$ for $-0.5 \leq \mu \leq 0.5$ (training) and $0.55 \leq \mu \leq 0.7$ (extrapolation).}
\label{re_im_eig2}
\end{figure}

The evolution of the eigenvalues versus $\mu$ is shown in Figs. \ref{re_im_eig1} and \ref{re_im_eig2} for the real and imaginary parts.  In Fig. \ref{re_im_eig1}, the real and imaginary parts of the first eigenvalue are shown.  This branch of eigenvalues corresponds to the oscillatory mode, $\lambda_1=\mu + 2\pi i$ in pre-bifurcation region. After bifurcation, the figure shows the eigenvalue that corresponds to the limit-cycle mode, $\lambda_1 = \frac{2\pi i}{1+\mu}$.   The real part moves in the positive direction at a rate of $1$ in the pre-bifurcation region and stays on the imaginary axis once it reaches the bifurcation point. Meanwhile, the imaginary part initially stays at $2\pi$ in the pre-bifurcation region, and when it reaches the bifurcation point, the frequency decreases with a form of $\frac{2\pi}{\mu+1}$. This is consistent with what the linearized system predicts in Eq. \eqref{eqn:lin_lco_case2}. Figure \ref{re_im_eig2} shows the real and imaginary parts of the purely decaying eigenvalue which is denoted as $\lambda_2$. For the pre-bifurcation region, it seems that the system identifies the $\lambda_{2,0}$ which is the addition of two principal eigenvalues in the pre-bifurcation region. Whereas for the post-bifurcation region, this is a principal eigenvalue at a decaying rate of  $-2$ whose imaginary part stays zero for the entire parameter space.  The eigenvalues are extrapolated using the trained EKBF model. For $\lambda_1$, both real and imaginary parts are predicted well at the considered parameter points. In the case of $\lambda_2$, the model underestimates the damping rate in the extrapolation regime while it accurately identifies zero imaginary part. 
Lastly, it is remarked that, in the training process, it is found that without lifting, the model fails to capture $\lambda_2$, highlighting the necessity of the lifting map in the observable.

Unlike the earlier case with monomial lifting, the eigenfunctions found by time-delayed embedding are not as clearly interpretable. This is because the observable consists of delayed states, which form a high-dimensional space. However, for this simple system, it is possible to visualize the property of eigenfunctions using three coordinates. In Figs. \ref{2d_delay_eigf_pre} and \ref{2d_delay_eigf_post}, the absolute value and phase of the two eigenfunctions are shown. The eigenfunctions are projected onto the three dimensional subspace spanned by $\{z(t), z(t+2\Delta t), z(t+4\Delta t)\}$. The points in the space are obtained from the trajectories that are used to train the EKBF model, and colored by the values of isochrons or isostables. For both eigenfunctions, the points that are further away from the center equilibrium have larger absolute value. One main difference between the two eigenfunction is that $\varphi_1$ has $360^{\circ}$ phase while $\varphi_2$ does not have such variation in the phase. % For $\varphi_2$, the values are on the negative real line, making the phase $\pi$ everywhere in the space.
This is because $\varphi_1$ corresponds to $\lambda_1 = \mu + 2 \pi i$ which provides a decaying oscillatory mode, whereas $\varphi_2$ corresponds to $\lambda_2= 2\mu$ which only provides a purely decaying mode with a fixed phase.

Figure \ref{2d_delay_eigf_post} shows the eigenfunctions after bifurcation. Figure \ref{2d_delay_eigf_post} (a) shows the eigenfunction corresponding to $\lambda_1 = 2\pi i/(1+\mu)$, meaning that only the phase information propagates forward in time along the limit cycle. Whereas the principal eigenfunction shown in Fig. \ref{2d_delay_eigf_post} (b) provides decaying mode, attracting the neighboring points to the limit cycle. The points that are further away from the limit cycle have stronger damping as they have larger absolute value. The phase information of $\varphi_2$ describes the direction of decay. The points outside of the limit cycle have phase of $\pi$, attracting them towards zero in positive direction, and the points inside of the limit cycle have phase of 0, attracting them towards zero in negative direction. As clearly shown in these figures, the eigenfunctions are still illustrative despite their high-dimensionality in the observable space.

\begin{figure}
\centering
\insertfigs{pics/2d_hopf_delay_eigenfunction_prebifurcation_phi1}{0.9}{Eigenfunction of the oscillatory mode, $\varphi_1$}
\insertfigs{pics/2d_hopf_delay_eigenfunction_prebifurcation_phi2}{0.9}{Eigenfunction of the purely decaying mode, $\varphi_2$}
\caption{Eigenfunctions defined in delayed coordinates at $\mu=-0.5$}
\label{2d_delay_eigf_pre}
\end{figure}

\begin{figure}
\centering
\insertfigs{pics/2d_hopf_delay_eigenfunction_postbifurcation_phi1}{0.9}{Eigenfunction of the oscillatory mode (limit-cycle mode), $\varphi_1$}
\insertfigs{pics/2d_hopf_delay_eigenfunction_postbifurcation_phi2}{0.9}{Eigenfunction of the purely decaying mode (Principal eigenfunction), $\varphi_2$}
\caption{Eigenfunctions defined in delayed coordinates at $\mu=0.5$.}
\label{2d_delay_eigf_post}
\end{figure}

\subsection{Nonlinear Panel Flutter}

Next, we apply the Koopman analysis to the supersonic panel flutter, a classical aeroelastic problem showing Hopf bifurcation \cite{Dowell1974}, as illustrated in Fig. \ref{cs2d}.  Consider a semi-infinite panel of length $L$ and linear piston theory, the dynamics is
\begin{equation}
    m\ddot{w} + (Nw_x)_x + D w_{xxxx} = \gamma p_\infty  (M_\infty w_x+ \dot{w}/a_\infty),
\end{equation}
where $m$ and $D$ are mass per length and bending stiffness, respectively, and $N=\int_0^L EI w_x^2 dx$ is the nonlinear in-plane force.
% $$
% N=\int_0^L EI w_x^2 dx.
% $$

\insertfighere{cs2d}{0.5}{Schematic of the 2D panel flutter problem \cite{Huang2018}.}

In its numerical solution, the problem is non-dimensionalized with flutter parameter $\Omega=\frac{\gamma p_\infty M_\infty^2 L^3}{D}$\footnote{In standard literature, e.g., \cite{Dowell1974}, the symbol for this parameter is usually $\lambda$.  But here $\Omega$ is used to avoid conflict with eigenvalues.}.
The parameters are chosen so that the flutter occurs at around $\Omega=517.4$.
The problem is solved using a nonlinear finite element method based on 3rd-order Hermite polynomials \cite{Huang2018}.
A total of 20 elements are used for discretization, but only the displacement, slope and their velocities at 3/4 chord are used as data for training, which provides 4 states out of 80 states in total.  This setup resembles the experimental conditions, where only limited measurements might be possible.

In the following, the EKBF model will be tested using both noiseless and noisy data for flutter analysis and prediction.  The Koopman eigenvalues are compared to those of the locally linearized models.
The method will be benchmarked against the standard autoregressive (AR) model \cite{box2015time}. The AR method has been used for decades to fit the time series data and flutter analysis.  In the following, the notation AR($N$) means an AR model fit using $N$ past measurements.  The metric of flutter margin \cite{zimmerman1964prediction} is used with the AR model for flutter prediction. % When the time response is captured well, the identified system approximates the spectrum of the true system sufficiently well and hence is appropriate to be used as a benchmark problem.

\subsubsection{Noiseless Data}

\begin{figure}[H]
\centering
\begin{minipage}{0.47\linewidth}
\begin{figure}[H]
\centering
\includegraphics[width=\textwidth]{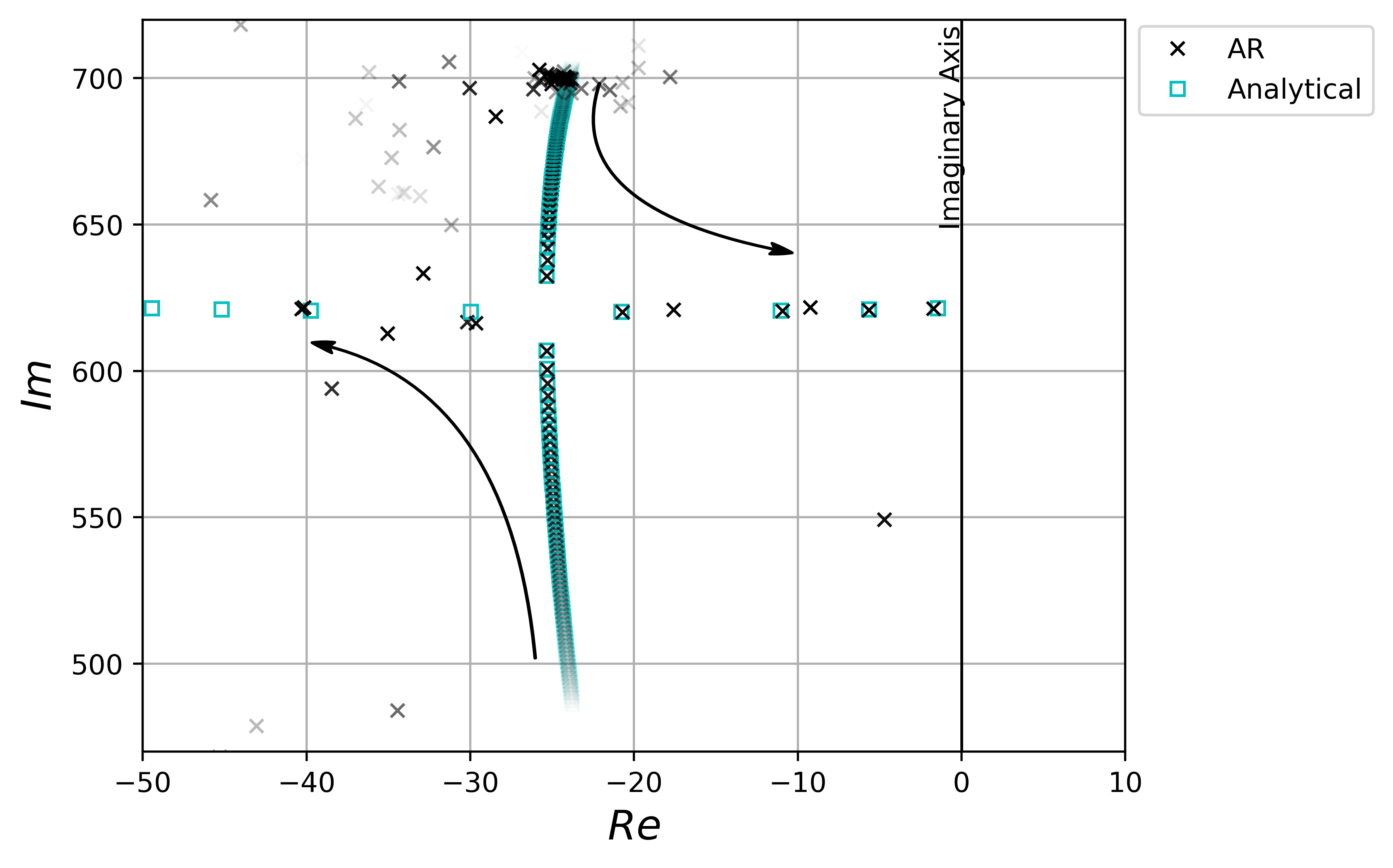}
\vspace{10pt}
\caption{Comparison of eigenvalues identified by AR and obtained by linearized model.}\label{noiseless_arma_eigs}
\end{figure}
\end{minipage}
\begin{minipage}{0.49\linewidth}
\begin{figure}[H]
\centering
\includegraphics[width=\textwidth]{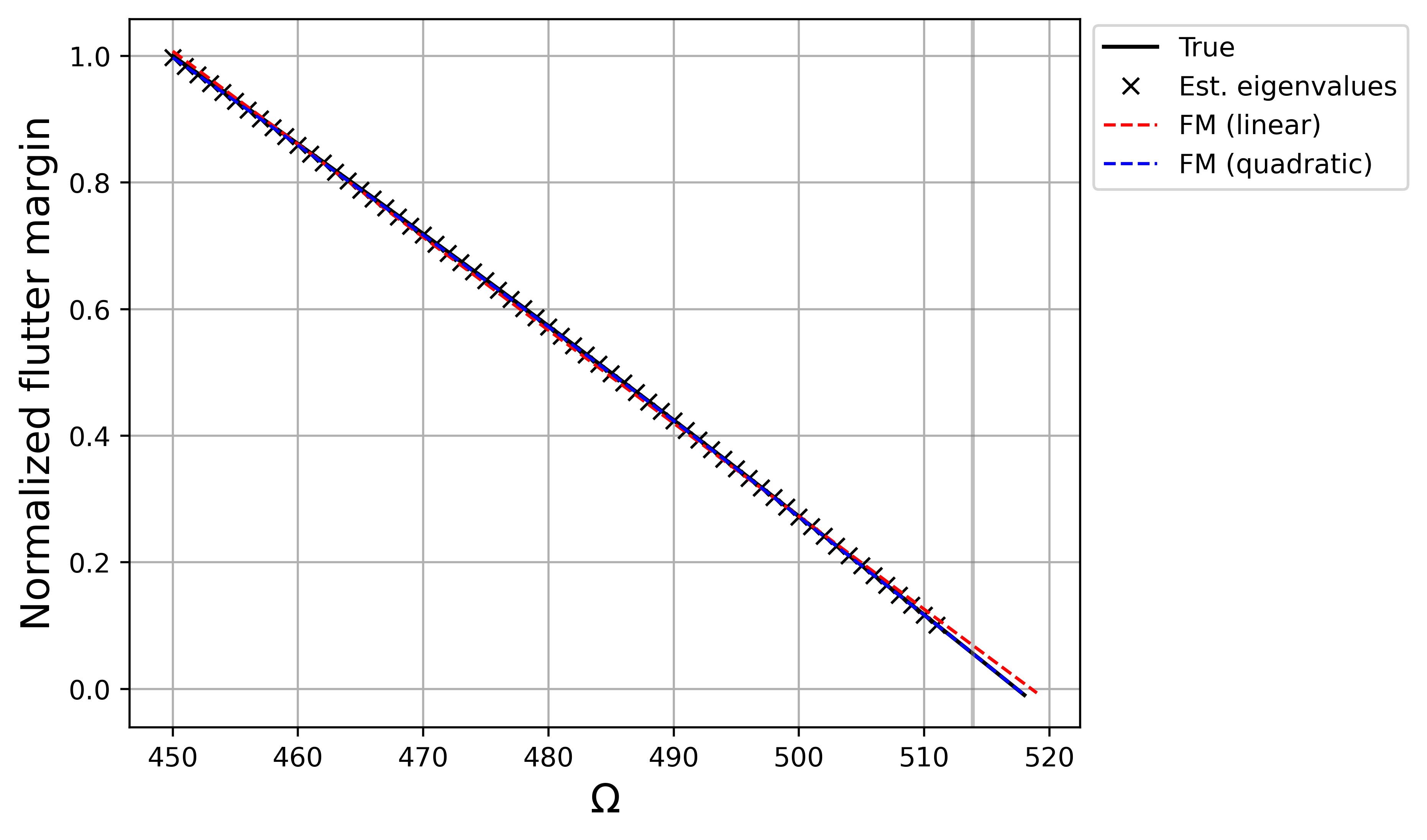}
\vspace{10pt}
\caption{Flutter prediction using flutter margin; calculated flutter margin is normalized by its maximum value for visualization purpose.}\label{noiseless_arma_fm}
\end{figure}
\end{minipage}
\end{figure}

To establish a baseline for subsequent comparison, Fig. \ref{noiseless_arma_eigs} shows the eigenvalues estimated using AR(80) model and the analytical eigenvalues (meaning the eigenvalues of the linearized FEM model). Due to the nonlinearity and high-dimensionality of the dynamics, the evolution of the eigenvalues are more complex than the previous 2D example problem. However, it clearly shows the flutter mechanism. The system initially starts with two discrete frequencies, and the two frequencies coalesce near the flutter. Then, one of the principal eigenvalues moves away from the imaginary axis, and the other one approaches to the imaginary axis. When this eigenvalue crosses over the imaginary axis, it indicates flutter of the system. It is important to note that the system exhibits hard flutter behavior where the eigenvalues of the system undergo rapid changes in their behavior near the flutter, making the prediction more challenging when pre-flutter data far from the flutter boundary are used.
In Fig. \ref{noiseless_arma_fm}, the flutter margin is calculated based on the analytical eigenvalues (denoted as True) and estimated eigenvalues from AR model (denoted as Est. eigenvalues).  The quadratic FM model provides very accurate estimation of the flutter boundary. The linear model predicts the flutter boundary to be $\Omega=518.6$ which is only $0.225 \%$ relative error. %The eigenvalues that are closest to the true values are used to calculate the flutter margin. In this case, there are evident branches that form frequency coalescence. Therefore, it is a practical approach that one can consider in a realistic scenario. 

% \begin{figure}[H]
%     \centering
%     \includegraphics[width=.8\textwidth]{pics/noiseless_arma_truth_comparison.png}
%     \caption{Comparison of eigenvalues identified by AR and obtained by linearized model.}\label{noiseless_arma_eigs}
% \end{figure}

% \begin{figure}[H]
%     \centering
%     \includegraphics[width=.7\textwidth]{pics/preflutter_coalescence.png}
%     \caption{Time response of panel flutter dynamics at $\Omega = 514$}\label{t_resp_coal}
% \end{figure}

% \begin{figure}[H]
% \centering
% \insertfigs{pics/ar_extrapolation}{0.8}{Real part of the eigenvalues from AR model and the extrapolation result for two polynomial extrapolations (P1 and P2); frequency coalescence is marked with vertical gray line}
% \insertfigs{pics/FM_noiseless}{0.8}{Flutter prediction using flutter margin; calculated flutter margin is normalized by its maximum value for visualization purpose}
% \caption{Real part of the eigenvalues from AR model and the extrapolation result for two polynomial extrapolations (P1 and P2)}
% \label{pf_arma_fm}
% \end{figure}

% \begin{center}
%     \insertfigr{noiseless_arma_truth_comparison}{0.45}{Comparison of eigenvalues identified by AR and obtained by linearized model.}
%     \insertfigr{FM_noiseless}{0.45}{Flutter prediction using flutter margin; calculated flutter margin is normalized by its maximum value for visualization purpose.}
% \end{center}

% \begin{figure}[H]
%     \centering
%     \includegraphics[width=.8\textwidth]{pics/FM_noiseless}
%     \caption{Flutter prediction using flutter margin; calculated flutter margin is normalized by its maximum value for visualization purpose.} \label{noiseless_arma_eigs}
% \end{figure}

In this experiment, we focus on the pre-flutter regime ($450\leq \Omega \leq 510$), and the prediction up to the flutter point is achieved by extrapolating the trained EKBF model. For each $\Omega$, 15 trajectories are sampled, with $\Omega$'s of increment 1.0. Each trajectory has a random initial condition that consists of the first two modes. The data are generated in such a way that approximately 7 time steps form one period for the lowest frequency of interest. A time delay embedding of 80 steps is used to train the 4th-order EKBF model. Example normalized trajectories for displacement and slope at 3/4 chord are shown in Fig. \ref{pf_ekbf_prediction}, demonstrating the accuracy of the identified EKBF model. For prediction, the trajectory before the vertical gray line is given to the EKBF model as it requires delayed coordinates to construct observable, and the model predicts the states after the gray line. The calculated normalized root-mean-square error (NRMSE) is low and on the order of $10^{-4}$.

\begin{figure}[H]
    \centering
    \includegraphics[width=.5\textwidth]{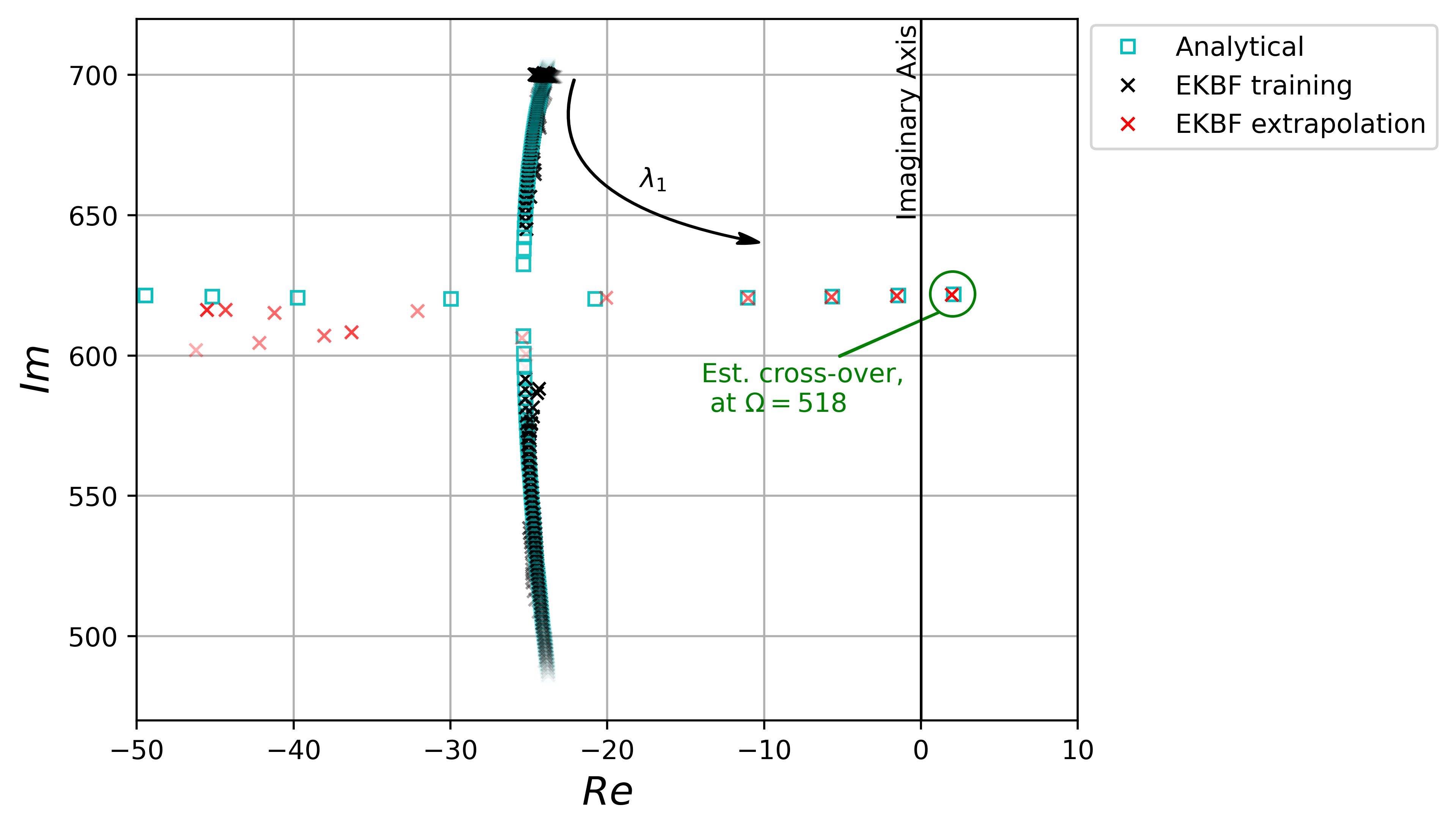}
    \caption{Analytical and estimated eigenvalues calculated by EKBF model for pre-flutter ($450\leq\Omega\leq 510$) and extrapolation ($511\leq \Omega \leq 518$), darker markers mean higher $\Omega$.}\label{flutter_eigs}
\end{figure}

The eigenvalues of the EKBF model are calculated for each parameter case considered in the experiment.  The root loci of the EKBF model are plotted in Fig. \ref{flutter_eigs} along with the analytical eigenvalues for comparison. The analytical result is obtained by directly solving the eigenvalue problem of FEM model around center equilibrium, so it cannot capture the limit cycle, which explains why it keeps moving towards the positive real part after flutter.  As mentioned in Sec. \ref{sec:num_alg}, when the least-squares is applied directly to obtain finite approximation of Koopman operator, it often results in numerous spurious eigenvalues even in the absence of noise. 
In the extrapolation phase, the MAC value of both right and left eigenvectors are calculated, and only the eigenvalues whose MAC values are greater than 0.89 are retained in the filtered eigenvalues. 
In Fig. \ref{flutter_eigs}, flutter prediction using EKBF model is presented. %The MAC threshold vlaue is set to be 0.93.
The extrapolated eigenvalues from EKBF model does not capture the exact analytical eigenvalues especially near the frequency coalescence due to the highly non-linear pattern. Nevertheless, it accurately predicts the eigenvalues near the flutter boundary.  Since the step size $\Delta \Omega=1$, it cannot predicts more accurate boundary unless the step size is refined. This is trivial to do, but once the cross-over at $\Omega=518$ is identified, one can simply apply linear interpolation between $\Omega=517$ and $\Omega=518$ to find the boundary. In this case, it is found to be $\Omega=517.4$, which is the true flutter boundary point.

% Note that the true flutter boundary is at $\Omega=517.4$, but the prediction can be refined by considering smaller parameter steps in the extrapolation process. This demonstrates the capability of EKBF model in both capturing the flutter mechanism and predicting the flutter boundary.

The two principal eigenfunctions at $\Omega=450$ are shown in Fig. \ref{pf_pre_eigf}. Since the EKBF model lives in 80 delayed coordinates space and due to its complex dynamics, it is difficult to visualize the calculated eigenfunctions. Therefore, after eigenfunctions are calculated, they are projected onto the linear subspace spanned by their first two principal directions using Principal Component Analysis (PCA) for better visualization. This means that each PCA dimension is a linear combination of 80 delayed coordinates used for EKBF model. Figure \ref{pf_pre_eigf} (a) shows the eigenfunction whose imaginary part of the eigenvalue (denoted as upper branch) is 704, and Fig. \ref{pf_pre_eigf} (b) shows the eigenfunction whose imaginary part of the eigenvalue is 485 (denoted as lower branch). In both branches, their phase forms $360^\circ$ of one cycle in the entire observable space. It is found that even though their real part of eigenvalues are approximately the same, the absolute value of eigenfunction is greater in the lower branch than the upper branch. This means that the lower branch decays faster than the upper branch, making the upper branch more sustaining oscillatory dynamics. In fact, this long-lasting oscillatory mode leads to limit cycle once the system enters flutter regime.

\begin{figure}[H]
\centering
\insertfigs{pics/pf_noiseless_ekbf_prediction_omega_480}{0.439}{$\Omega=480$}
\insertfigs{pics/pf_noiseless_ekbf_prediction_omega_510}{0.546}{$\Omega=510$}
\caption{Normalized EKBF model time response prediction from unseen initial conditions.}
\label{pf_ekbf_prediction}
\end{figure}

\begin{figure}[H]
\centering
\insertfigs{pics/pf_eig_function_preflutter_eig_idx_0_phi1}{0.8}{Upper branch at $\Omega= 450$, \;\; $\varphi_1$}
\insertfigs{pics/pf_eig_function_preflutter_eig_idx_0_phi2}{0.8}{Lower branch at $\Omega=450$, \;\; $\varphi_2$}
\caption{Principal eigenfunctions of panel flutter dynamics for pre-flutter case.}
\label{pf_pre_eigf}
\end{figure}

\subsubsection{Noisy Data}

\begin{figure}[H]
\centering
\begin{minipage}{0.47\linewidth}
\begin{figure}[H]
\centering
\includegraphics[width=\textwidth]{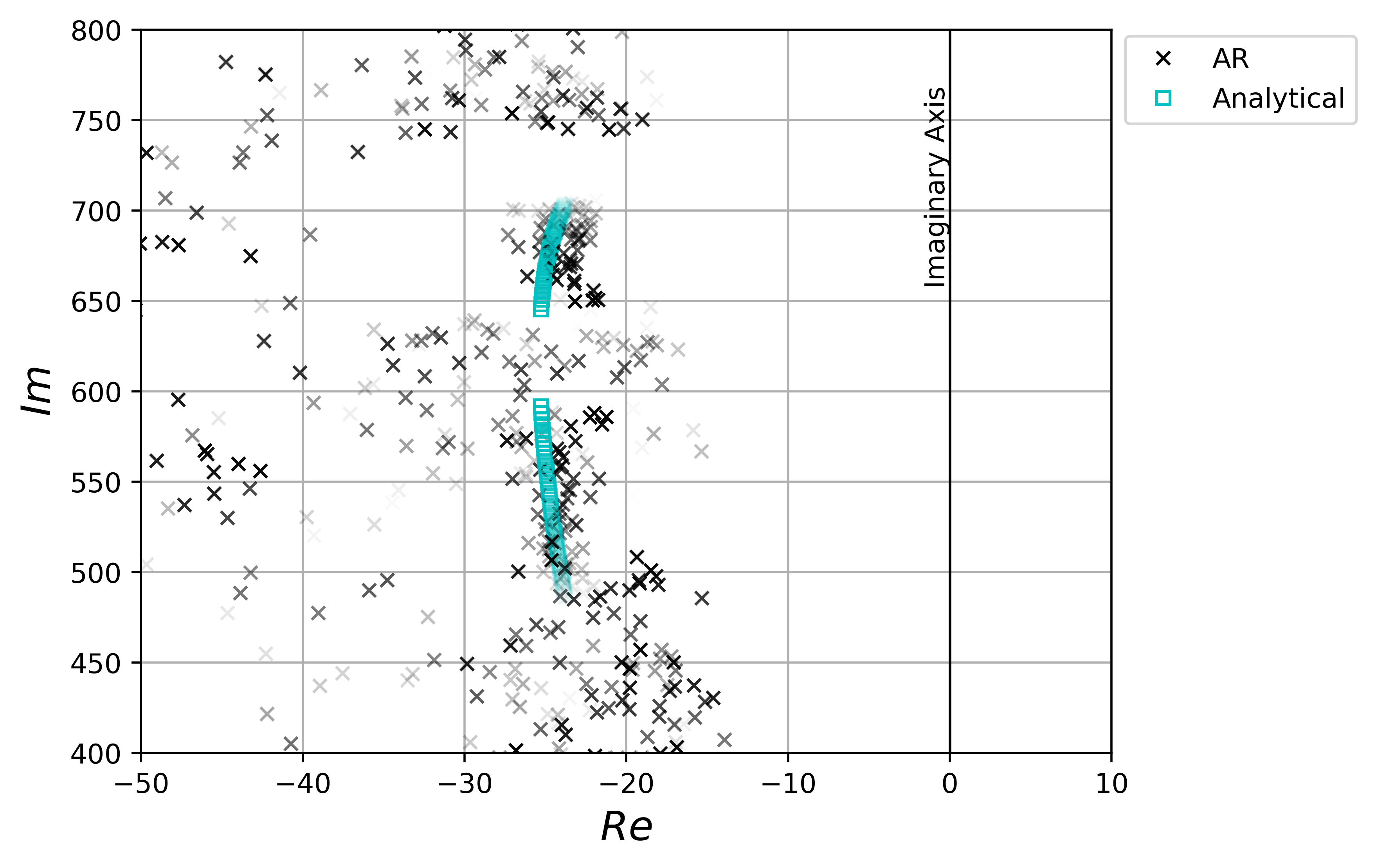}
\vspace{10pt}
\caption{Eigenvalues calculated by AR model using noisy measurement for pre-flutter regime ($ 450 \leq \Omega \leq 510$).}\label{pf_ar_noise_eigs}
\end{figure}
\end{minipage}
\begin{minipage}{0.49\linewidth}
\begin{figure}[H]
\centering
\includegraphics[width=\textwidth]{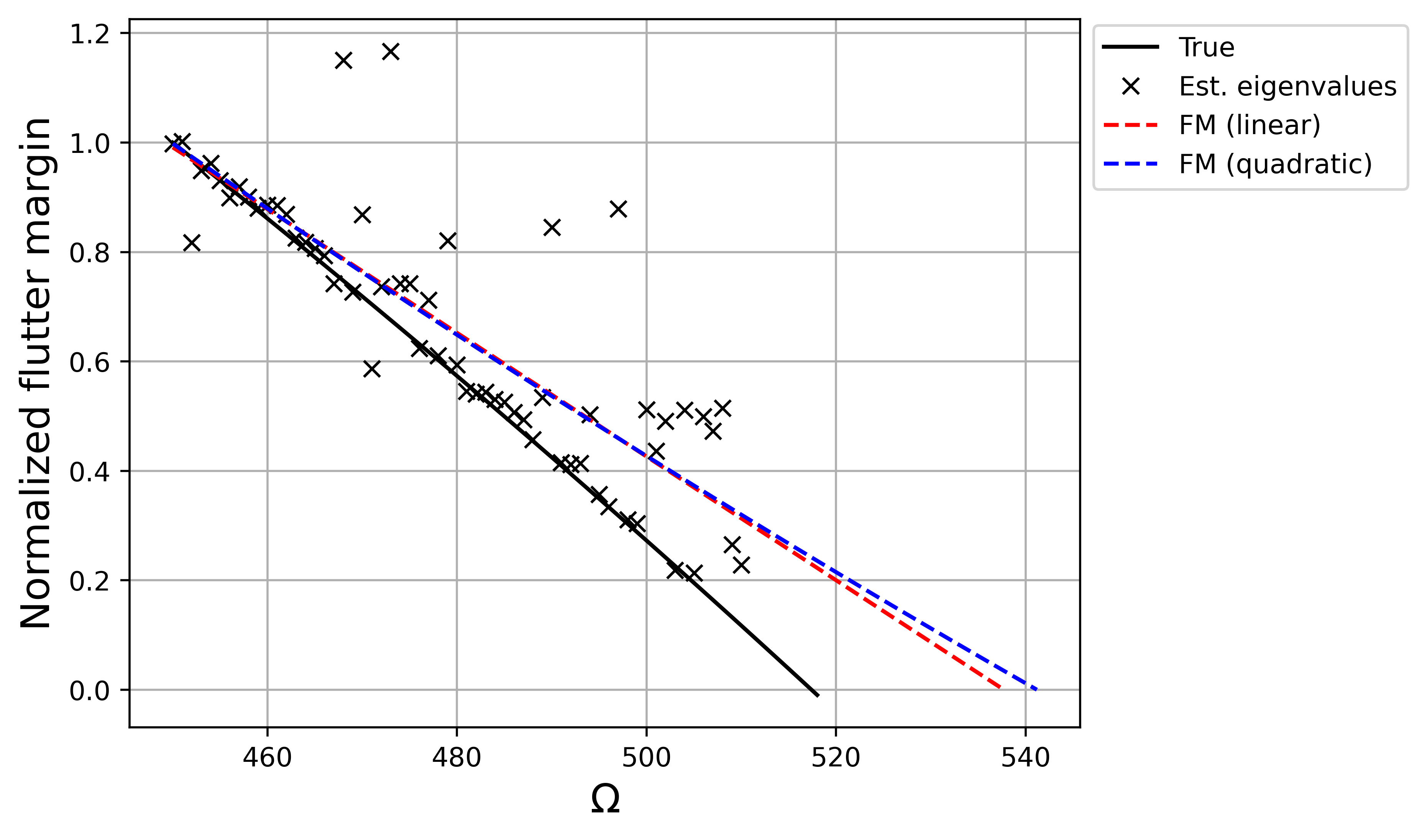}
\vspace{10pt}
\caption{Computed flutter margin along with the linear and quadratic interpolators.}\label{pf_noise_fm}
\end{figure}
\end{minipage}
\end{figure}

In this section, we consider noisy measurement of panel flutter dynamics. We particularly focus on the state-independent measurement noise. This means for every parameter $\Omega$, the noise is added such that
\begin{equation}
    \tilde{\vx}(t) = \vx(t) + \vu(t),
\end{equation}
where $\vu(t) \sim \mathcal{N}(\mathbf{0}, \vtS^2)$ is a zero-mean multi-variate Gaussian noise with $\vtS^2 = \text{diag}(\sigma_1^2, \sigma_2^2, \dots , \sigma_n^2)$, and the variance values $\sigma_j^2$ are fixed for all $\Omega$. In Fig. \ref{pf_ar_noise_eigs}, calculated eigenvalues using AR (80) model is shown. Using 80 delayed states, AR model can successfully predict the future states. Nevertheless, the eigenvalues are polluted by the noise in the data and high-dimensional delay embedding in the observable. In the figure, without analytical eigenvalues, it is difficult to see the right branch of principal eigenvalues. As a result, selecting appropriate eigenvalue pair to calculate flutter margin is infeasible in this case. One relatively simple approach would be to apply Fourier transform of the noisy data to identify dominant frequencies and find the eigenvalues that have closest imaginary part to the identified frequencies. In Fig. \ref{pf_noise_fm}, estimated flutter margin based on the eigenvalues using Fourier transform is shown along with both linear and quadratic trend line. Contrast to the noiseless case, calculated flutter margin values spread out due to the noise. As a result, the predicted flutter boundary by the trend line deviated from the truth significantly.  The linear FM model predicts the flutter boundary to be 537.7, which is 3.92 \% relative error while the quadratic FM model predicts the flutter boundary to be 541.2 which is 4.60 \% relative error, and it is even worse than the linear model prediction result.  The discrepancy between the two trend lines becomes more significant than the noiseless case. This means that the prediction of flutter boundary can depend on the choice of extrapolation method. 

\begin{figure}[H]
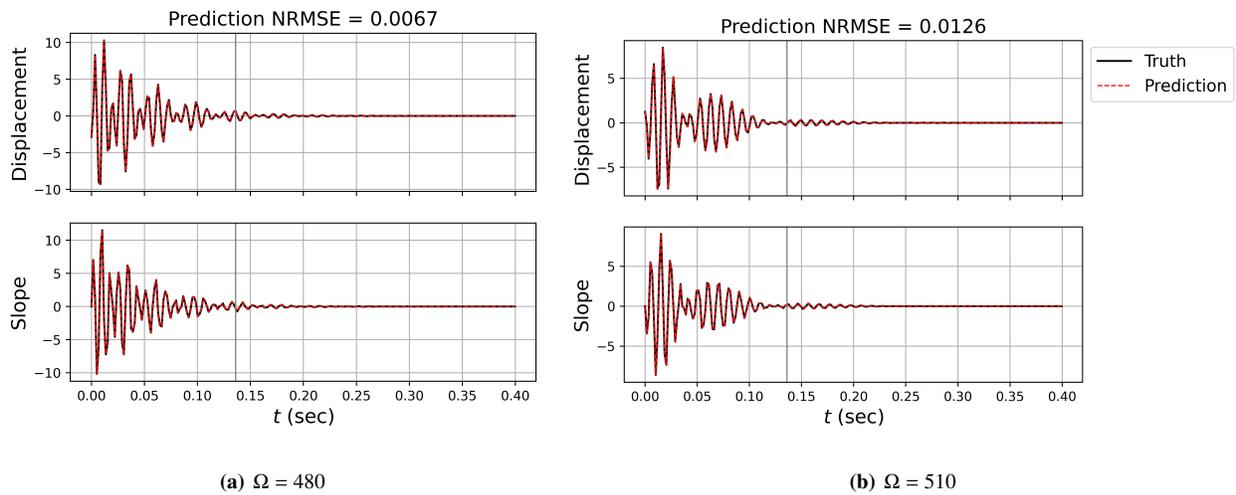

\centering
\insertfigs{pics/pf_noise_ekbf_prediction_omega_480}{0.439}{$\Omega=480$}
\insertfigs{pics/pf_noise_ekbf_prediction_omega_510}{0.546}{$\Omega=510$}
\caption{EKBF model time response prediction from unseen initial condition.}
\label{pf_ekbf_noise_prediction}
\end{figure}

Meanwhile, the flutter prediction using EKBF model does not depend on the extrapolation method, as the parametric nature of EKBF model captures the right dependency on the parameters in the system.  The accuracy of the EKBF model is demonstrated by the prediction of the time response as shown in Fig. \ref{pf_ekbf_noise_prediction}.  The model can successfully predict the future states given some unseen initial trajectories; %since the model is defined in delayed coordinates, it requires a set of delayed points as initial condition.
like before, the prediction is done after the vertical gray line due to the delayed coordinates.  Therefore, simple parameter sweep over the extrapolation regime can provide accurate flutter prediction. Figure \ref{pf_ekbf_noise_eigs} shows the EKBF eigenvalues for both training and extrapolation regime using noisy measurement. The MAC threshold value is set to be 0.9.  Despite slight error in the frequency, the EKBF model captures correct eigenvalue motion after training regime, accurately predicting the flutter boundary at $\Omega=518$. For more accurate flutter boundary estimation, a simple linear interpolation is done between $\Omega=517$ and 518.  The interpolation provides the flutter boundary at $\Omega=517.4$ which is the true flutter boundary point. 

% \begin{figure}[H]
%     \centering
%     \includegraphics[width=.8\textwidth]{pics/pf_ar_noise_eigs.png}
%     \caption{Eigenvalues calculated by AR model using noisy measurement for pre-flutter regime ($ 450 \leq \Omega \leq 510$)}\label{pf_ar_noise_eigs}
% \end{figure}

% \begin{figure}[H]
%     \centering
%     \includegraphics[width=.8\textwidth]{pics/noise_FM.png}
%     \caption{Computed flutter margin along with the linear and quadratic interpolators}\label{pf_noise_fm}
% \end{figure}

\begin{figure}[H]
\centering
\insertfigs{pics/pf_noise_eigs_61}{0.9}{Complex plane}
\insertfigs{pics/noise_flutter_prediction_ekbf}{0.915}{Real part of eigenvalues for different system parameter, $\Omega$}
\caption{Eigenvalue comparison for both training and extrapolation regime.}
\label{pf_ekbf_noise_eigs}
\end{figure}

\subsubsection{EKBF Model Using Both Pre-flutter and Post-flutter Data}

In the previous section, we presented flutter prediction using pre-flutter data, leveraging the system identification of the EKBF model.  The parametric form of EKBF model allows continuity in the variation of eigenfunction. In this section, we evaluate the capability of EKBF in capturing both pre-flutter and post-flutter data.

\begin{figure}[H]
    \centering\includegraphics[width=.8\textwidth]{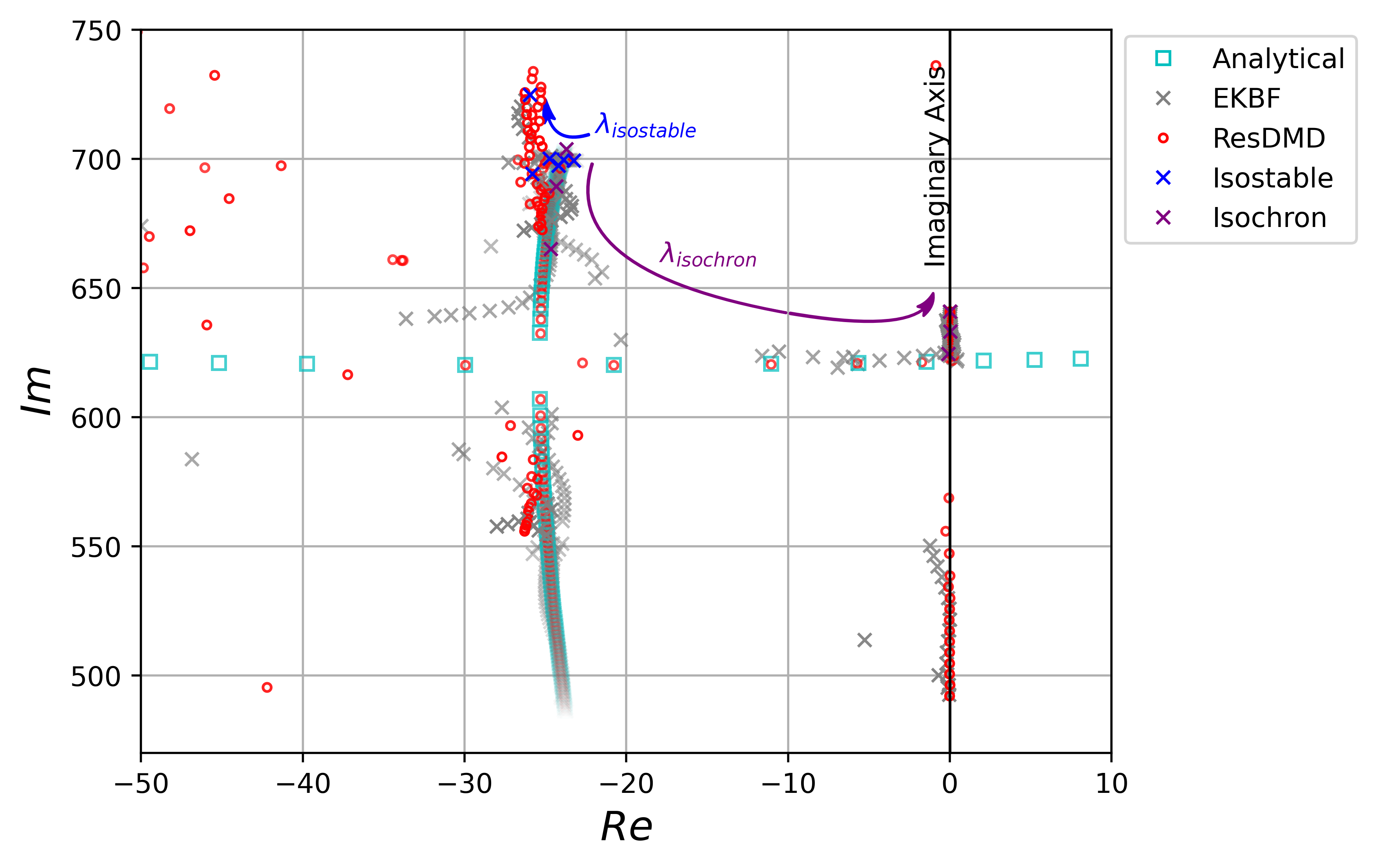}
    \caption{Analytical and estimated eigenvalues calculated by EKBF model for both pre-flutter and post-flutter ($450 \leq \Omega \leq 540$).}\label{pre_post_flutter_eigs}
\end{figure}

In Fig. \ref{pre_post_flutter_eigs}, the eigenvalues obtained by analytical method, EKBF, and ResDMD methods are presented. For ResDMD, a separate model is trained for each flutter parameter, so there is no parameterization involved. The ResDMD result shows a good agreement with the analytical one up to the pre-flutter regime, demonstrating its accuracy. After the flutter boundary, ResDMD still captures the limit cycle frequencies on the imaginary axis. It is found that the identified EKBF model can capture this behavior, with two dominant families of eigenvalues. One family of eigenvalues, denoted $\lambda_{isostable}$, are almost stationary around $Im(\lambda) = 700$ and move upward after flutter.  Note that the ResDMD filtering threshold number needs to increase to capture this family of eigenvalues, which is why more eigenvalues have appeared in the plot.  The other family of eigenvalues, denoted $\lambda_{isochron}$, are initially in the negative real part and lie on the imaginary axis after flutter.

% In Fig. \ref{pre_post_flutter_eigs}, the eigenvalues for isostable are denoted by blue cross mark, and those for isochron are denoted by purple cross mark.

\begin{figure}[H]
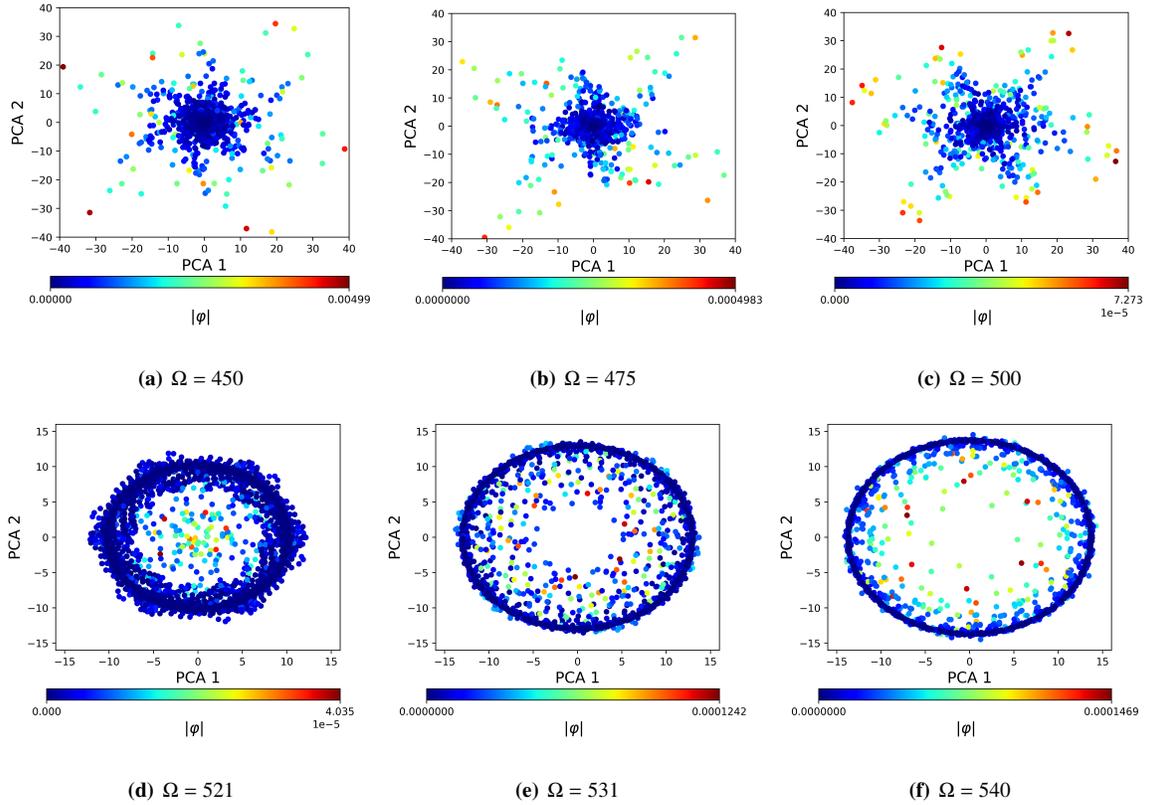

\centering
\insertfigs{pics/pf_isostable_preflutter_eig_idx_0}{0.3}{$\Omega = 450$}
\insertfigs{pics/pf_isostable_preflutter_eig_idx_25}{0.3}{$\Omega = 475$}
\insertfigs{pics/pf_isostable_preflutter_eig_idx_50}{0.29}{$\Omega = 500$}
\insertfigs{pics/pf_isostable_postflutter_eig_idx_71}{0.29}{$\Omega = 521$}
\insertfigs{pics/pf_isostable_postflutter_eig_idx_81}{0.3}{$\Omega = 531$}
\insertfigs{pics/pf_isostable_postflutter_eig_idx_90}{0.3}{$\Omega = 540$}
\caption{EKBF isostable corresponding to $\lambda_{isostable}$ of panel flutter system.}
\label{pf_prepost_isostable}
\end{figure}

\begin{figure}[H]
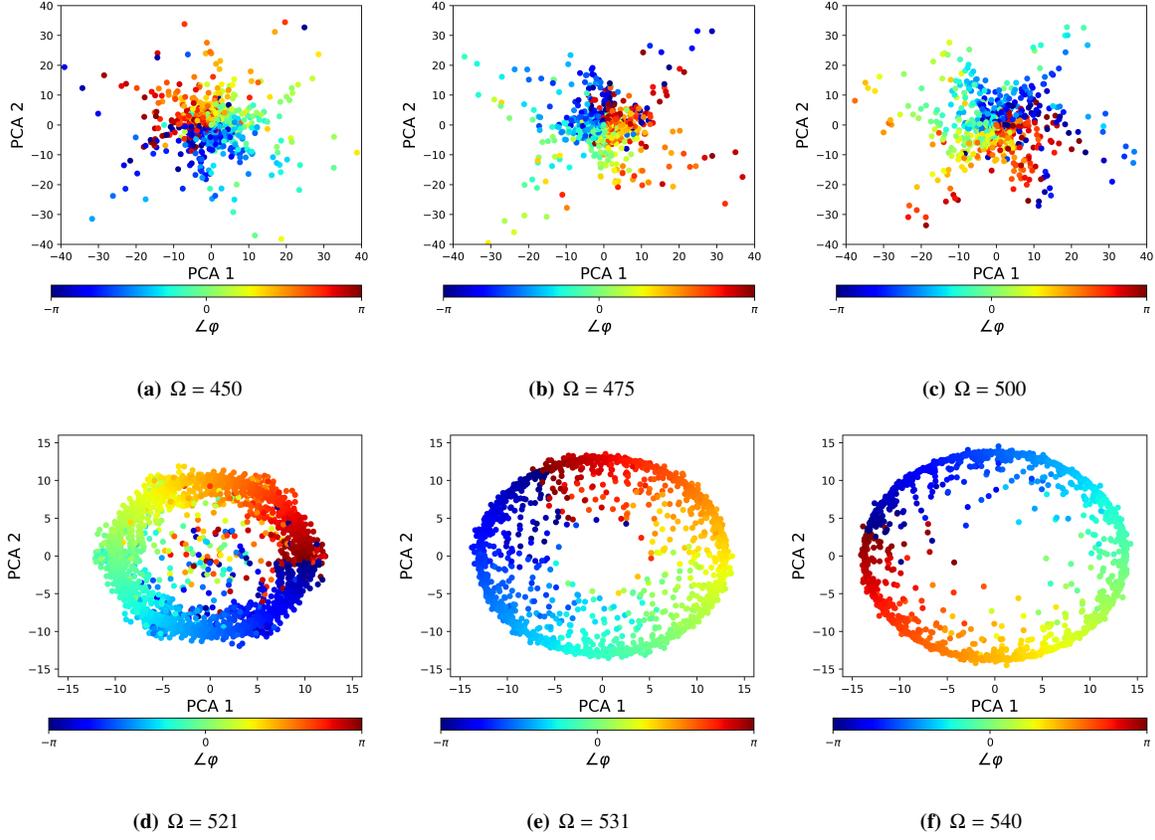

\centering
\insertfigs{pics/pf_eig_function_preflutter_eig_idx_0}{0.3}{$\Omega = 450$}
\insertfigs{pics/pf_eig_function_preflutter_eig_idx_25}{0.3}{$\Omega = 475$}
\insertfigs{pics/pf_eig_function_preflutter_eig_idx_50}{0.3}{$\Omega = 500$}
\insertfigs{pics/pf_eig_function_postflutter_eig_idx_71}{0.3}{$\Omega = 521$}
\insertfigs{pics/pf_eig_function_postflutter_eig_idx_81}{0.3}{$\Omega = 531$}
\insertfigs{pics/pf_eig_function_postflutter_eig_idx_90}{0.3}{$\Omega = 540$}
\caption{EKBF isochron corresponding to $\lambda_{isochron}$ of panel flutter system.}
\label{pf_prepost_isocrhon}
\end{figure}

To examine the characteristics of the two families, their eigenfunctions are shown in Fig. \ref{pf_prepost_isostable} and \ref{pf_prepost_isocrhon}, respectively.
Figure \ref{pf_prepost_isostable} shows the change in isostable corresponding to $\lambda_{isostable}$ at 6 different system parameters, whose eigenvalues are represented by blue cross mark in Fig. \ref{pre_post_flutter_eigs}; note that these eigenpairs are calculated from the same EKBF model.  This branch of modes provides modal damping even after the flutter occurs and their frequencies are similar to the frequency of the flutter, and hence is deemed appropriate to discuss its isostable. The first three eigenfunctions are for pre-flutter and the second three eigenfunctions are for post-flutter.  One can see that in pre-flutter regime, it captures the property of the isostable that the points further away from the center fixed points have stronger damping, and the points near the center fixed point are closer to zero.  Once the flutter occurs, the system forms limit cycle, and the points further away from the limit cycle have stronger damping. The initial conditions of the system are mostly less than the amplitude of the limit cycle, which makes the cluster inside of the limit cycle and not the outside.

Figure \ref{pf_prepost_isocrhon} shows the change in isochron corresponding to $\lambda_{isochron}$ at 6 different system parameters whose eigenvalues are represented by purple cross mark in Fig. \ref{pre_post_flutter_eigs}.  Note that the panel system is complex considering the number of states in the finite element model. As a result, in pre-flutter case, the space in which the function lives is complicated. Therefore, the phase is not as clear as simple 2D case we considered before.  Once the system enters post-flutter regime, limit cycle is formed and all the neighboring points are attracted to this new basin of attraction. Now, due to the existence of the limit-cycle, the underlying structure of the system becomes rather simpler.  This is because the system trajectories are defined in the neighborhood of the limit cycle space, reducing the complexity of the space.  As one can see in Fig. \ref{pf_prepost_isocrhon} (d)-(f), when PCA is applied to the eigenfunction, their phase is much cleaner than in the preflutter case. 

% For pre-flutter case, all the points in the system asymptotically decay to the global attracting point.

% \begin{figure}[H]
%     \centering
%     \includegraphics[width=.8\textwidth]{pics/eigs_pre_post.png}
%     \caption{Analytical and estimated eigenvalues calculated by EKBF model for both pre-flutter and post-flutter ($450 \leq \Omega \leq 540$) \hl{30 eigenvalues are retained}}\label{pre_post_flutter_eigs}
% \end{figure}

% \begin{figure}[H]
%     \centering
%     \includegraphics[width=.8\textwidth]{pics/eigs_pre_post2.png}
%     \caption{Analytical and estimated eigenvalues calculated by EKBF model for both pre-flutter and post-flutter ($450 \leq \Omega \leq 540$) \hl{40 eigenvalues are retained, you can see more eigenvalues going upwards after flutter}}\label{pre_post_flutter_eigs}
% \end{figure}

% \hl{After flutter occurs, there are many eigenvalues that have negative real part (even outside the scope of the provided figure), but I think the ones that have similar frequency of LCO, are more important ones. You can see there is one branch that are going upwards after flutter near $690 < Im(\lambda) < 730$. I think this is the one I need to show for isostable plot}

% \todo[linecolor=red,backgroundcolor=red!25,bordercolor=red]{I wanted to put colors on the eigenvalues I selected for the plot below, but the preflutter eigenvalues will be hidden by others because they are too compact.}

\section{Conclusions}\label{sec3}

In this paper, we presented a parametrization method of an aeroelastic system based on the Koopman theory. We especially utilized existing parameterized Koopman model, so-called Koopman bilinear form. To overcome the limitation of the linearity in KBF model, we developed Extended KBF (EKBF) model to account for stronger nonlinear dependence on, e.g., the flutter parameter. The global linearization endowed by EKBF provides globally linear system describing the aeroelastic dynamics.  The eigensystem of the global linear system is connected to those of the nonlinear system, thus extending the classical (local) linear stability analysis for flutter to a global linear stability analysis, that can be directly applied to the analysis of limit cycles.

Specifically, we show that the principal Koopman eigenvalues generalize the linear eigenvalues in the fixed point case, and the Floquet eigenvalues in the limit cycle case.  Due to the exact correspondence, the nonlinear system stability can be determined using the same criteria in the classical linearized cases (i.e., the signs of real parts).
In addition, the Koopman eigenfunctions allow the computation of isostables and isochrons with relative ease.  The isostable, as the magnitude of an eigenfunction, reveals the Lyapunov stability of the nonlinear dynamics.  The isochron, as the phase of an eigenfunction, shows the synchronization characteristics of trajectories starting from different initial conditions, which can be further leveraged for, e.g., phase reduction analysis.

Based on the EKBF model, a flutter analysis and prediction framework is developed, which consists of the design of lifting functions, the least-squares solution for the EKBF model, and two filtering techniques, ResDMD and dual MAC, for the mitigation of spurious eigenpairs.
The ResDMD is useful in the diagnostic case, i.e., the analysis of existing data, by selecting the most plausible eigenpairs produced by the numerical solver.
The dual MAC is developed to tackle the predictive case, i.e., the extrapolation of the model to flutter boundary, by selecting eigenpairs that are more similar to those obtained from data.

The proposed methods are demonstrated on a 2D academic example and a more realistic panel flutter problem.  In the academic example, we derived the analytical form of Poincaré map to motivate the necessity of EKBF model. We showed that trained EKBF model can interpolate and even extrapolate the principal eigenvalues in the unseen parameter regime.  We also showed the eigenfunctions defined in the delayed coordinates to provide the interpretation of the eigenfunction embedded in high-dimensional space.  In the panel flutter problem, the Koopman eigenvalues were compared to the eigenvalues of linearized finite element model to show the correspondence between them. We further compared eigenvalue predictions from the autoregressive (AR) model, highlighting the EKBF model's superior capability in extrapolating flutter boundary, even in the presence of measurement noise. Unlike traditional approaches that are sensitive to noise, the parametric nature of EKBF allows reliable flutter prediction through systematic parameter sweeps. By taking advantage of principal component analysis (PCA) for eigenfunction interpretation, this method offers deeper insights into the phase and stability characteristics of panel flutter dynamics.

Nevertheless, there is still room for improvement. Especially, the estimated eigenvalues by EKBF is not sufficiently accurate when the range of parameter is broad involving both pre-flutter and post-flutter regimes. More importantly, the current framework only estimates the flutter boundary if only the pre-flutter data is available. In general, one may also seek the entire bifurcation diagram, and uncover the type of bifurcation, e.g. supercritical or subcritical. The above issue may be solved by a closer exploration of the connection between the methods based on critical slowing down (CSD) and Koopman theory through EKBF. %, the current methodology could be improved to identify the entire bifurcation diagram.

\section*{Acknowledgments}

D.H. acknowledges the support from the NSF CAREER program CMMI-2340266.

\bibliography{references}

\end{document}